\documentclass[10pt,twocolumn,english,prd,superscriptaddress,nofootinbib,preprintnumbers,showpacs,floatfix]{revtex4-2}

\usepackage{graphicx,color}
\usepackage{amsmath}
\usepackage{amssymb}
\usepackage{hyperref}
\usepackage[utf8]{inputenc}
\usepackage[english]{babel}
\usepackage{epsfig}
\usepackage{subfigure}
\usepackage{wasysym}
\usepackage{color,xcolor}
\usepackage{amsmath}
\usepackage{bm}
\usepackage{epsfig}
\usepackage{amsfonts}
\usepackage{dcolumn}
\usepackage{float}

\usepackage{makecell} 
\usepackage{comment}

\usepackage{cases}

\hypersetup{colorlinks=true, linkcolor=blue, citecolor=green}

\usepackage{dcolumn}
\usepackage{bm}
\usepackage{ifpdf}
\usepackage{hyperref}
\usepackage{float}
\usepackage{bm}
\usepackage{xcolor,color,graphicx,graphics}
\usepackage[OT1]{fontenc}
\usepackage{latexsym,amssymb,amsmath,amsfonts}
\usepackage{makeidx}
\usepackage{epsfig}
\usepackage{epstopdf}
\usepackage{mathrsfs}
\hypersetup{colorlinks=true, linkcolor=blue, citecolor=green}
\usepackage{enumerate}
\usepackage{xcolor}
 \usepackage{multirow}

\usepackage{tikz}

\newcommand{\orcidicon}{%
	\begin{tikzpicture}
	\draw[lime, fill=lime] (0,0) 
		circle [radius=0.16] 
		node[white] {{\fontfamily{qag}\selectfont \tiny ID}};
	\draw[white, fill=white] (-0.0625,0.095) 
		circle [radius=0.007];
	\end{tikzpicture}	\hspace{-2mm}
}
\newcommand\orcidEdnaldo{{\href{https://orcid.org/0000-0002-9388-8373}{\orcidicon}}}
\newcommand\orcidFrancisco{{\href{https://orcid.org/0000-0002-9388-8373}{\orcidicon}}}
\newcommand\orcidManuel{{\href{https://orcid.org/0000-0001-8586-0285}{\orcidicon}}}
\newcommand\orcidTarciso{{\href{https://orcid.org/0009-0007-0450-2672}{\orcidicon}}}
\newcommand\orcidHenrique{{\href{https://orcid.org/0000-0001-7565-4277}{\orcidicon}}}
\newcommand\orcidLuis{{\href{https://orcid.org/0009-0009-4322-6484}{\orcidicon}}}
\newcommand\orcidJorde{{\href{https://orcid.org/0009-0001-3344-2986}{\orcidicon}}}

\newcommand\orcidOlmo{{\href{https://orcid.org/0000-0001-9857-0412}{\orcidicon}}}
\begin{document}
%

\title{Black bounce sourced by non-minimally coupled linear electrodynamics and a canonical scalar field through a thin shell at the throat}

\author{Ednaldo L. B. Junior\orcidEdnaldo\!\!} \email{ednaldobarrosjr@gmail.com}
\affiliation{Faculdade de F\'{i}sica, Universidade Federal do Pará, Campus Universitário de Tucuruí, CEP: 68464-000, Tucuruí, Pará, Brazil}
\affiliation{Programa de P\'{o}s-Gradua\c{c}\~{a}o em F\'{i}sica, Universidade Federal do Sul e Sudeste do Par\'{a}, 68500-000, Marab\'{a}, Par\'{a}, Brazill}

	\author{Jos\'{e} Tarciso S. S. Junior\orcidTarciso\!\!}
 \email{tarcisojunior17@gmail.com}
\affiliation{Faculdade de F\'{\i}sica, Programa de P\'{o}s-Gradua\c{c}\~{a}o em 
F\'isica, Universidade Federal do 
 Par\'{a},  66075-110, Bel\'{e}m, Par\'{a}, Brazil}

	\author{Francisco S. N. Lobo\orcidFrancisco\!\!} \email{fslobo@ciencias.ulisboa.pt}
\affiliation{Instituto de Astrof\'{i}sica e Ci\^{e}ncias do Espa\c{c}o, Faculdade de Ci\^{e}ncias da Universidade de Lisboa, Edifício C8, Campo Grande, P-1749-016 Lisbon, Portugal}
\affiliation{Departamento de F\'{i}sica, Faculdade de Ci\^{e}ncias da Universidade de Lisboa, Edif\'{i}cio C8, Campo Grande, P-1749-016 Lisbon, Portugal}

\author{Gonzalo J. Olmo\orcidOlmo\!\!}
  \email{gonzalo.olmo@uv.es}
\affiliation{Instituto de F\'isica Corpuscular (IFIC), CSIC‐Universitat de Val\`{e}ncia, Spain}
\affiliation{Universidade Federal do Cear\'a (UFC), Departamento de F\'isica,\\ Campus do Pici, Fortaleza - CE, C.P. 6030, 60455-760 - Brazil.}

	\author{Jorde A. A. Ramos\orcidJorde\!\!}
 \email{jordealves@ufpa.br}
\affiliation{Faculdade de F\'{\i}sica, Programa de P\'{o}s-Gradua\c{c}\~{a}o em 
F\'isica, Universidade Federal do 
 Par\'{a},  66075-110, Bel\'{e}m, Par\'{a}, Brazil}
 
	\author{Manuel E. Rodrigues\orcidManuel\!\!}
	\email{esialg@gmail.com}
	\affiliation{Faculdade de F\'{\i}sica, Programa de P\'{o}s-Gradua\c{c}\~{a}o em 
F\'isica, Universidade Federal do 
 Par\'{a},  66075-110, Bel\'{e}m, Par\'{a}, Brazil}
\affiliation{Faculdade de Ci\^{e}ncias Exatas e Tecnologia, 
Universidade Federal do Par\'{a}\\
Campus Universit\'{a}rio de Abaetetuba, 68440-000, Abaetetuba, Par\'{a}, 
Brazil}


\author{Luís F. Dias da Silva\orcidLuis\!\!} 
        \email{fc53497@alunos.fc.ul.pt}
\affiliation{Instituto de Astrof\'{i}sica e Ci\^{e}ncias do Espa\c{c}o, Faculdade de Ci\^{e}ncias da Universidade de Lisboa, Edifício C8, Campo Grande, P-1749-016 Lisbon, Portugal}


 \author{Henrique A. Vieira\orcidHenrique\!\!} \email{henriquefisica2017@gmail.com}
\affiliation{Faculdade de F\'{i}sica, Programa de P\'{o}s-Gradua\c{c}\~{a}o em F\'{i}sica, Universidade Federal do Par\'{a}, 66075-110, Bel\'{e}m, Par\'{a}, Brazil}

\date{\LaTeX-ed \today}

\begin{abstract}
 We construct a novel class of black bounce solutions within General Relativity, sourced by a canonical scalar field non-minimally coupled to linear electrodynamics, establishing a self-consistent framework in which regular black holes and traversable wormholes are supported by ordinary bulk matter, with the necessary exoticity confined to an infinitesimally thin defect at the bounce.
\end{abstract}

\maketitle
\def\HMS{{\scriptscriptstyle{\rm HMS}}}

\section{Introduction}\label{sec1}

Despite the remarkable observational progress achieved in the last decade, such as gravitational wave detections and event horizon imaging~\cite{Einstein:1916vd,LIGOScientific:2016aoc,LIGOScientific:2017ync,EventHorizonTelescope:2019dse,EventHorizonTelescope:2022wkp}, which back up General Relativity (GR) as a successful phenomenological description of nature, the theory is still believed to be inherently incomplete due to the inevitable emergence of spacetime singularities in processes of gravitational collapse. To resolve the problems associated with singular boundaries, extensive effort has been devoted to constructing regular, singularity-free geometries. A prominent line of research in this direction concerns regular black holes (RBHs)~\cite{Ansoldi:2008jw}, pioneered by Bardeen's ad-hoc metric, which describes a Reissner-Nordstrom like object with horizons but with a core of finite curvature~\cite{Bardeen:1968}. Ayón-Beato and García showed that Bardeen-like spacetimes can be rigorously embedded as exact solutions of GR coupled to non-linear electrodynamics (NEDs), with the regularization parameter acting as a magnetic monopole charge~\cite{Ayon-Beato:2000mjt, Rodrigues:2018bdc}. Such configurations tend to regularize the inner region via a de Sitter or Minkowski core~\cite{Culetu:2015cna, Simpson:2019mud, Bronnikov:2024izh}. Historically, NEDs were introduced by Born and Infeld~\cite{Born:1933pep, Born:1934gh} to get rid of  point-charge self-energy divergences in classical Maxwell theory. Since then, a plethora of NED Lagrangians have been considered in multiple and diverse gravitational contexts~\cite{Heisenberg:1936nmg,Plebanski:1966,Kruglov:2014hpa, Kruglov:2014iwa, Kruglov:2014iqa, Bandos:2020jsw, Harko:2013xma, Harko:2013wka, Harko:2013aya}.

Minimally coupling NED to GR has yielded a rich landscape of RBH models~\cite{Bronnikov:2000vy, Dymnikova:2004zc, Balart:2014cga, Culetu:2014lca} with varied physical imprints. For instance, effective photon trajectories are modified by non-linear field effects~\cite{Novello:1999pg, Habibina:2020msd, Toshmatov:2021fgm, dePaula:2024yzy}, which distorts black hole shadows~\cite{Stuchlik:2019uvf, Allahyari:2019jqz, Kruglov:2020tes}. Standard thermodynamic relations are also altered~\cite{Breton:2004qa, Myung:2007xd, Ma:2015gpa, Kruglov:2016ymq, Fan:2016hvf}. Extended frameworks within modified gravity theories also heavily employ NED to generate non-singular geometries~\cite{Junior:2015fya, Rodrigues:2015ayd, Rodrigues:2016fym, deSousaSilva:2018kkt, Rodrigues:2019xrc, Tangphati:2023xnw, Junior:2023ixh, Junior:2024xmm} (for a broad overview, see Ref.~\cite{Lan:2023cvz}).

An alternative mechanism to cure black hole singularities was proposed by Simpson and Visser~\cite{Simpson:2018tsi} (expanding on earlier concepts in Ref.~\cite{Visser:1997yn}) under the name of  \textit{black bounce} (BB). The BB mechanism proposes to resolve central singularities by replacing the radial area coordinate via $r \to \sqrt{r^2 + a^2}$, ensuring a non-vanishing minimal area as $r\to 0$. The bounce parameter $a$ allows the metric to smoothly interpolate between standard/regular black holes and traversable wormholes. Though wormholes were once regardered as just useful educational tools for teaching GR,  traversable wormhole physics has grown over the years into a major domain of study~\cite{PhysRev.48.73,Bronnikov:1973fh, Ellis:1973yv, Morris:1988cz, Barcelo:1999hq, Barcelo:2000zf, Visser:2003yf, Lobo:2005us, Lobo:2007zb, Cardoso:2016rao, Bronnikov:2017sgg, Lobo:2017cay, Blazquez-Salcedo:2020czn, Churilova:2021tgn, Konoplya:2021hsm}.

The BB paradigm has sparked extensive research into energy conditions and causal structures~\cite{Lobo:2020ffi}, traversable geometry bounds~\cite{Lobo:2020kxn, Berry:2020tky}, gravitational lensing patterns~\cite{Nascimento:2020ime, Tsukamoto:2020bjm, Cheng:2021hoc, Tsukamoto:2021caq, Zhang:2022nnj}, and potential observational signatures~\cite{Guerrero:2021ues, Jafarzade:2021umv, Jafarzade:2020ova, Jafarzade:2020ilt, Yang:2021cvh, Bambhaniya:2021ugr, Ou:2021efv, Guo:2021wid, Wu:2022eiv, Tsukamoto:2022vkt}. Extensions to stationary rotating metrics~\cite{Muniz:2024wiv,Mazza:2021rgq, Xu:2021lff}, string cloud sources~\cite{Rodrigues:2022rfj, Yang:2022ryf}, and cylindrical symmetries~\cite{Bronnikov:2023aya, Lima:2022pvc} have further generalized the setup. While many BB models rely on NED or coupled scalar-NED fields with magnetic/electric/dyonic charges~\cite{Bronnikov:2022bud, Canate:2022gpy, Rodrigues2023, Pereira:2023lck, Lima:2023arg, Alencar:2024yvh, Junior:2025sjr}, alternative constructions within modified gravity theories are also widely documented~\cite{Olmo:2012nx,Junior:2022zxo, Junior:2023qaq, Junior:2024vrv, Junior:2024cbb, Rois:2024qzm, Rois:2025czu}. A general metric-engineering formalism for static, spherically symmetric BB spacetimes was recently formulated in Ref.~\cite{Alencar:2025jvl}.

Despite the widespread use of NED to build non-singular spacetimes, this paradigm suffers from non-trivial theoretical and physical drawbacks. In fact,  many classic NED-supported geometries (including Bardeen's model) fail to recover the standard Maxwell linear limit in the weak-field asymptotic regime, which is in clear conflict with elementary physics~\cite{Mignani:2016fwz, Ejlli:2020yhk}. On the other hand,  optical trajectories in NED are governed by an effective spacetime metric~\cite{Novello:1999pg} and modified energy-transport equations ~\cite{DeFelice:2026nti}, which may affect key geometric properties (such as regularity) and complicate the analysis of observational features like shadows~\cite{daSilva:2023jxa}. Moreover, physically acceptable NED theories should obey fundamental criteria such as causality (subluminal signal propagation) and unitarity~\cite{Shabad:2011hf}. However, regular magnetic configurations generically violate these principles near their centers. On the field-theory side, the electrodynamic Lagrangians $\mathcal{L}(F)$ required for electric RBH/BB solutions are frequently non-analytic or multivalued~\cite{Bronnikov:2000vy, Rodrigues:2018bdc, Alencar:2024yvh}, raising questions about their physical reality. Additionally, NED configurations are often subject to linear dynamic instabilities~\cite{Moreno:2002gg, Breton:2014nba, Toshmatov:2019gxg, Nomura:2020tpc, DeFelice:2024seu, DeFelice:2024ops}. These difficulties suggest that formulations rooted in linear electrodynamics (LED) could offer a significantly more transparent and self-consistent alternative.

Inspired by the methodology proposed in Ref.~\cite{RBHLE} for regular black holes, here we present a framework for generating non-singular spacetimes within GR using linear electrodynamics with a  non-minimal coupling to a scalar field of the form $W(\varphi)\mathcal{L}(F)$. This setup allows for the analytical reconstruction of the supporting matter content and recovers the standard minimally coupled scalar-electrodynamics when $W(\varphi) = 1$. In this paper, we extend this non-minimal coupling method to black-bounce geometries, going beyond the Bardeen-type backgrounds studied in Ref.~\cite{RBHLE}. We determine the functional forms of $W(\varphi)$ and $\mathcal{L}(F)$ required to sustain regular geometries but requiring that the Maxwellian behavior be recovered in the weak-field regime. This provides a simpler, pathology-free route to modelling non-singular configurations such as traversable wormholes. In this process, we will see that physically consistent configurations that minimize the violation of the energy conditions can indeed be obtained. 

The remainder of this manuscript is organized as follows. In Section~\ref{sec2}, we outline the field equations for static, spherically symmetric spacetimes with non-minimally coupled scalar and electrodynamic fields, focusing on magnetically charged configurations. After deriving the field equations, we interpret them in terms of anisotropic fluids, which facilitates the discussion of the energy conditions of the model presented in Section~\ref{sec:ModeI}. Here we check the regularity of the model and reconstruct the matter Lagrangian, coupling function, scalar field profile, and scalar potential. Section \ref{sec:matching} is devoted to the careful analysis of the radial bounce surface, where a thin shell that violates the energy conditions is found. Finally, Section~\ref{sec:concl} summarizes our findings. We adopt geometrized units ($G = c = 1$) and the metric signature $(+,-,-,-)$.

\section{Field equations coupled to nonlinear electrodynamics and scalar field}\label{sec2}

\subsection{Action and field equations}

To begin, let us consider the Einstein-Hilbert gravitational action,
a scalar field, and a non-minimal interaction term that couples the
scalar and electromagnetic sectors, as shown
below~\cite{Cordeiro:2025ivw}
\begin{equation}
	S = \int \sqrt{-g} \, d^4x \left[ R
	- 2\kappa^2\big( \mathcal{L}_\varphi(\varphi)
	- \mathcal{L}_I (\varphi,F)\big) \right],
	\label{action}
\end{equation}
where $\kappa^2 = 8\pi$.  The Lagrangian of the scalar field is
\begin{equation}
	\mathcal{L}_\varphi(\varphi)
	= \epsilon(\varphi)\,\partial^\mu\varphi\,\partial_\mu\varphi
	- V(\varphi),
	\label{Lphi}
\end{equation}
with
\begin{equation}
	\epsilon(\varphi)
	\begin{cases}
		>0, & \text{canonical scalar field},\\[4pt]
		<0, & \text{phantom scalar field},
	\end{cases}
\end{equation}
and $V(\varphi)$ denotes the scalar field potential.

The Lagrangian $\mathcal{L}_I(\varphi,F)$ in the
action~\eqref{action} describes the interaction between the scalar
field and the electromagnetic sector.  In this work, we adopt the
following explicit form for this term:
\begin{equation}
	\mathcal{L}_I(\varphi,F) = W(\varphi)\,\mathcal{L}(F).
	\label{LI}
\end{equation}
Here $W(\varphi)$ encodes the non-minimal coupling between the
electromagnetic and scalar fields, and $\mathcal{L}(F)$ is the
Lagrangian density of nonlinear electrodynamics (NED), which
depends on the electromagnetic invariant
\begin{equation}
	F = \frac{1}{4}\,F^{\mu\nu}F_{\mu\nu},
\end{equation}
where the electromagnetic field tensor is defined in terms of the
vector potential $A_\mu$ as
$F_{\mu\nu} = \partial_\mu A_\nu - \partial_\nu A_\mu$.
For the particular case $W(\varphi) = 1$, the standard NED scenario
with minimal coupling is recovered.

Varying the action~\eqref{action} with respect to the potential
$A_\mu$ and the scalar field $\varphi$ yields the following equations
of motion:
\begin{equation}
	\nabla_\mu \big( W(\varphi)\,\mathcal{L}_F(F)\,F^{\mu\nu} \big)
	= 0,
	\label{sol2}
\end{equation}
and
\begin{align}
	2\epsilon(\varphi)\,\nabla_\mu\nabla^\mu\varphi
	+ \nabla_\mu\varphi\,\nabla^\mu\varphi\,
	\frac{d\epsilon(\varphi)}{d\varphi}
	+ \mathcal{L}(F)\,\frac{dW(\varphi)}{d\varphi}
    &\nonumber\\
	= -\frac{dV(\varphi)}{d\varphi}&,
	\label{sol3}
\end{align}
where we denote $\mathcal{L}_F = \partial\mathcal{L}(F)/\partial F$.

Varying the action~\eqref{action} with respect to the metric tensor
$g_{\mu\nu}$, we obtain the gravitational field equation
\begin{equation}
	G^{\mu}_{\;\nu}
	\equiv R^{\mu}_{\;\nu} - \frac{1}{2}\,\delta^{\mu}_{\;\nu}R
	= \kappa^{2}\,T^{\mu}_{\;\nu}
	= \kappa^{2}\left( W(\varphi)\,\overset{F}{T}{}^{\mu}_{\;\nu}
	+ \overset{\varphi}{T}{}^{\mu}_{\;\nu}
	\right),
	\label{EqM}
\end{equation}
where $G^{\mu}_{\;\nu}$ is the Einstein tensor,
$\overset{F}{T}{}^{\mu}_{\;\nu}$ is the stress-energy tensor
associated with the NED, and $\overset{\varphi}{T}{}^{\mu}_{\;\nu}$
is the stress-energy tensor of the scalar field.

These stress-energy tensors are given, respectively, by
\begin{equation}
	\overset{F}{T}{}^{\mu}_{\;\nu}
	= \delta^{\mu}_{\;\nu}\,\mathcal{L}(F)
	- \mathcal{L}_{F}\,F^{\mu\alpha}F_{\nu\alpha},
	\label{TNED}
\end{equation}
and
\begin{equation}
	\overset{\varphi}{T}{}^{\mu}_{\;\nu}
	= \epsilon(\varphi)\,
	\bigl( 2\,\partial^{\mu}\varphi\,\partial_{\nu}\varphi
	- \delta^{\mu}_{\;\nu}\,
	\partial^{\sigma}\varphi\,\partial_{\sigma}\varphi
	\bigr)
	+ \delta^{\mu}_{\;\nu}\,V(\varphi).
	\label{Tphi}
\end{equation}

In the following, we present the line element used in our
investigation, together with the algebraic framework employed to
derive the solutions.

\subsection{Spherically symmetric black bounce}
\label{sec:spherically_symmetric}

To develop our solutions, we consider the following static and
spherically symmetric metric
\begin{equation}
	ds^2 = A(r)\,dt^2 - B(r)\,dr^2
	- \Sigma(r)^2 \bigl( d\theta^2 + \sin^2\theta\,d\phi^2 \bigr),
	\label{m}
\end{equation}
where $A(r)$ and $B(r)$ are functions of the radial coordinate $r$,
and $\Sigma(r)$ is the area function, which implements the bounce
in the radial coordinate.

For our purposes, we are interested in developing only solutions
with magnetic charge.  Thus, the only non-zero component of the
Maxwell--Faraday tensor $F^{\mu\nu}$ is
\begin{equation}
	F^{23} = \frac{q_m \csc\theta}{\Sigma^4(r)},
	\label{23}
\end{equation}
where $q_m$ represents the magnetic charge.  The electromagnetic
scalar $F$ then takes the form
\begin{equation}
	F = \frac{q_m^2}{2\,\Sigma^4(r)}.
	\label{F}
\end{equation}

To check the consistency of the solutions found, we will use the
identity
\begin{equation}
	\mathcal{L}_F - \frac{\partial\mathcal{L}}{\partial r}
	\left( \frac{\partial F}{\partial r} \right)^{-1} = 0,
	\label{RC}
\end{equation}
which follows directly from the chain rule.

With these considerations, the tensors~\eqref{TNED}
and~\eqref{Tphi} take the diagonal forms
\begin{equation}
	\overset{F}{T}{}^{\mu}_{\;\nu}
	= \mathcal{L}\;
	\mathrm{diag}\left( 1,\, 1,\,
	1 - \frac{q_m^2\,\mathcal{L}_F}{\Sigma^4\,\mathcal{L}},\,
	1 - \frac{q_m^2\,\mathcal{L}_F}{\Sigma^4\,\mathcal{L}}
	\right),
	\label{TNED2}
\end{equation}
and
\begin{equation}
	\overset{\varphi}{T}{}^{\mu}_{\;\nu}
	= \frac{\epsilon\,\varphi'(r)^2}{B(r)}\;
	\mathrm{diag}\bigl( 1,\, -1,\, 1,\, 1 \bigr)
	+ \delta^{\mu}_{\;\nu}\,V(\varphi).
	\label{Tphi2}
\end{equation}

\subsubsection{Magnetic solutions}

We now present the general solutions obtained by considering only
the magnetic charge component in the Maxwell--Faraday tensor, as
expressed in Eq.~\eqref{23}.  Substituting this into the Einstein
field equations~\eqref{EqM}, together with the metric~\eqref{m}
and Eqs.~\eqref{23} and~\eqref{F}, we obtain the following
components:
\begin{align}
	G_{\phantom{0}0}^{0}=\kappa^{2}T_{\phantom{0}0}^{0}=&\kappa^{2}\left(\frac{\epsilon(r)\varphi'(r)^{2}}{B(r)}+W(r){\cal L}(r)+V(r)\right),\label{EqF00}
	\\
	G_{\phantom{0}1}^{1}=\kappa^{2}T_{\phantom{0}1}^{1}=&\kappa^{2}\left(-\frac{\epsilon(r)\varphi'(r)^{2}}{B(r)}+W(r){\cal L}(r)+V(r)\right),\label{EqF11}
	\\
	G_{\phantom{0}2}^{2}=	\kappa^{2}T_{\phantom{0}2}^{2}=&\kappa^{2}\Bigg[W(r)\left({\cal L}(r)-\frac{q_m^{2}{\cal L}_{F}(r)}{\Sigma(r)^{4}}\right)
	\nonumber
	\\
	&
	+
	\frac{\epsilon(r)\varphi'(r)^{2}}{B(r)}+V(r)
	\Bigg]\,.\label{EqF22}
\end{align}

From Eqs.~\eqref{EqF00} and~\eqref{EqF22}, we can determine general
expressions for the NED Lagrangian $\mathcal{L}(r)$ and its
derivative $\mathcal{L}_F(r)$, given by
\begin{align}
	& {\cal L}(r) =\Big\{ B(r)^{2}\Big(1-\kappa^{2}\Sigma(r)^{2}V(r)\Big)
	\nonumber
	\\
	& -B(r)\Big[\Sigma'(r)^{2}+\Sigma(r)\Big(2\Sigma''(r)+\kappa^{2}\Sigma(r)\epsilon(r)\varphi'(r)^{2}\Big)\Big]
	\nonumber
	\\
	&    +\Sigma(r)B'(r)\Sigma'(r)\Big\}\Big/\left(\kappa^{2}W(r)\Sigma(r)^{2}B(r)^{2}\right)\, ,	\label{L_BB}
\end{align}
and
\begin{eqnarray}
	&&    {\cal L}_{F}(r)=\frac{\Sigma(r)^{2}}{4\kappa^{2}q_m^{2}A(r)^{2}B(r)^{2}W(r)}\Big\{-B(r)\Sigma(r)^{2}A'(r)^{2}
	\nonumber
	\\
	&& \qquad
	+2A(r)^{2}\big[-2B(r)\left(\Sigma(r)\Sigma''(r)+\Sigma'(r)^{2}\right)+2B(r)^{2}
	\nonumber
	\\
	&& \qquad
	+\Sigma(r)B'(r)\Sigma'(r)\big]+A(r)\Sigma(r)\big[2B(r)\Sigma(r)A''(r)
	\nonumber
	\\
	&& \qquad
	+A'(r)\left(2B(r)\Sigma'(r)-\Sigma(r)B'(r)\right)\big]\Big\}\,	,\label{LF_BB}
\end{eqnarray}
respectively.

We observe that Eqs.~\eqref{L_BB} and~\eqref{LF_BB} satisfy the
components~\eqref{EqF00} and~\eqref{EqF22} of Einstein's field
equations.  Equation~\eqref{EqF11} provides an additional relation
that allows us to determine $\epsilon(r)$:
\begin{eqnarray}
	&& \frac{2 A(r) B(r) \Sigma ''(r)-\Sigma '(r) \left(B(r) A'(r)+A(r) B'(r)\right)}{A(r) B(r)^2 \Sigma (r)}
	\nonumber\\
	&& \hspace{4cm}  +\frac{2 \kappa ^2 \epsilon (r) \varphi '(r)^2}{B(r)}=0 \,.\label{comp00}
\end{eqnarray}
From Eq.~\eqref{comp00}, we determine the parameter $\epsilon(r)$ as

\begin{align}
 	\epsilon(r)	&=\frac{\Sigma'(r)\Big(B(r)A'(r)+A(r)B'(r)\Big)}{2\kappa^{2}A(r)B(r)\Sigma(r)\varphi'(r)^{2}}
	-\frac{\Sigma''(r)}{\kappa^{2}\Sigma(r)\varphi'(r)^{2}}.\label{eps}
\end{align}

To obtain the potential $V(r)$, we employ the equation of
motion~\eqref{sol3}.  Substituting $\epsilon(r)$ from
Eq.~\eqref{eps} and performing a direct integration, we arrive at
the following general expression:
\begin{widetext}
	\begin{eqnarray} 
		V(r) &=& W(r)\int\frac{1}{2\kappa^{2}W(r)^{2}A(r)\Sigma(r)^{2}B(r)^{3}}\Bigg\{ B(r)^{2}\Big[\Sigma'(r)\Big[\Sigma(r)W(r)A''(r)+A'(r)\Big(\Sigma(r)W'(r)+3W(r)\Sigma'(r)\Big)\Big]\Big]
		\nonumber\\&&
		\qquad \quad -B(r)^{2}\Sigma(r)W(r)A'(r)\Sigma''(r)+A(r)\Bigg[-2\Sigma(r)W(r)B'(r)^{2}\Sigma'(r)+B(r)\Bigg(3\Sigma(r)W(r)B'(r)\Sigma''(r)
		\nonumber\\&&
		\qquad \quad  -2B(r)^{2}W'(r)
		+\Sigma'(r)\Big[\Sigma(r)W(r)B''(r)+B'(r)\Big(3W(r)\Sigma'(r)-\Sigma(r)W'(r)\Big)\Big]
		\nonumber\\&&
		\qquad \quad +2B(r)\Big[W'(r)\Big(\Sigma(r)\Sigma''(r)+\Sigma'(r)^{2}\Big)
		-W(r)\left(\Sigma(r)\Sigma'''(r)+3\Sigma'(r)\Sigma''(r)\Big)\right]\Bigg)\Bigg]\Bigg\}dr
		.\label{V}
	\end{eqnarray}
	
	After defining $V(r)$ as described above, we adopt an approach to
	simplify the solutions developed later.  To this end, we set the
	derivative of the Lagrangian given by Eq.~\eqref{LF_BB} to
	${\cal L}_F = 1$.  Consequently, we can determine the general
	expression for $W(r)$,
	\begin{align}
		W(r)=&\frac{\Sigma(r)^{2}}{4\kappa^{2}q_{m}^{2}A(r)^{2}B(r)^{2}}\Big\{-B(r)\Sigma(r)^{2}A'(r)^{2}+A(r)\Sigma(r)\left[2B(r)\Sigma(r)A''(r)+A'(r)\left(2B(r)\Sigma'(r)-\Sigma(r)B'(r)\right)\right]
		\nonumber\\&  +2A(r)^{2}\left[\Sigma(r)B'(r)\Sigma'(r)-2B(r)\left(\Sigma(r)\Sigma''(r)+\Sigma'(r)^{2}\right)+2B(r)^{2}\right]\Big\}\label{W}.
	\end{align}
\end{widetext}
Its explicit form in terms of the radial coordinate depends solely
on the choice of the metric functions and the area function, as
shown by Eq.~\eqref{W}.

It should be noted that, after assuming ${\cal L}_F = 1$ in
Eq.~\eqref{LF_BB}---regardless of any specific assignment for the
metric functions $A(r)$, $B(r)$, the area function $\Sigma(r)$,
and $\varphi(r)$---this choice necessarily corresponds to the
linear case, i.e., linear electrodynamics (LED) with
$\mathcal{L} \sim F$.
Therefore, with the choice given in Eq.~\eqref{W},
Eq.~\eqref{L_BB} becomes
\begin{equation}
	{\cal L}  (r)= \frac{q_m^2}{2 \Sigma (r)^4}.\label{L}
\end{equation}
Its explicit form depends solely on the choice of the area function.
Subsequently, when we address the specific model chosen for the
metric function $A(r)$, we shall appropriately introduce the
function $\Sigma(r)$.

Thus, we have completed all the equations required for the general
analysis of the system with a non-minimal coupling between the
scalar field and, following the definition in~\eqref{L}, the LED.
From these expressions, it suffices to specify functional forms for
the metric functions $A(r)$ and $B(r)$, the area function
$\Sigma(r)$, and the scalar field $\varphi(r)$ in order to obtain
explicit solutions.  In the next section, we present a model
resulting from a specific choice of these functions.

\subsection{Matter described by an anisotropic fluid}

It was shown in Ref.~\cite{Lessa:2024erf} that general
energy--momentum tensors associated with matter sources modelled by
NED and a scalar field, within a static and spherically symmetric
configuration, can be interpreted as an equivalent anisotropic fluid.
To begin our discussion, we first introduce the energy--momentum
tensor of an anisotropic fluid:
\begin{equation}
	T_{\mu\nu} = (\rho + p_t)u_\mu u_\nu - p_t g_{\mu\nu}
	+ (p_r - p_t)v_\mu v_\nu,
	\label{TME_anis}
\end{equation}
where $u_\mu$ is the timelike four-velocity, $v_\mu$ is the spacelike
four-velocity in the radial direction, $\rho$ is the energy density,
$p_r = p_r(r)$ the radial pressure, and $p_t = p_t(r)$ the tangential
pressure.  This fluid is characterised by the following mixed
components:
\begin{equation}
	T^{\mu}_{\;\nu} = \text{diag}\left( \rho,\, -p_{r},\, -p_{t},\, -p_{t} \right).
	\label{Comp_TA}
\end{equation}

Adopting the line element~\eqref{m} and treating the matter sector
as an anisotropic fluid in the field equations~\eqref{EqM}, one
obtains the following expressions:
\begin{align}
	\kappa^2\rho(r) &=
	\frac{B'(r)\Sigma'(r)}{B(r)^2\Sigma(r)}
	- \frac{2\Sigma''(r)}{B(r)\Sigma(r)}
	- \frac{\Sigma'(r)^2}{B(r)\Sigma(r)^2}
	\nonumber\\ &
    + \frac{1}{\Sigma(r)^2}; \label{rho} \\[4pt]
	\kappa^2p_r(r) &=
	\frac{A'(r)\Sigma'(r)}{A(r)B(r)\Sigma(r)}
	+ \frac{\Sigma'(r)^2}{B(r)\Sigma(r)^2}
	- \frac{1}{\Sigma(r)^2}; \label{pr} \\[4pt]
	\kappa^2p_t(r) &=
	\frac{A'(r)\Sigma'(r)}{2A(r)B(r)\Sigma(r)}
	- \frac{A'(r)B'(r)}{4A(r)B(r)^2}
	+ \frac{\Sigma''(r)}{B(r)\Sigma(r)}
		\nonumber \\
		&\quad
	+ \frac{A''(r)}{2A(r)B(r)} 
	- \frac{B'(r)\Sigma'(r)}{2B(r)^2\Sigma(r)}
	- \frac{A'(r)^2}{4A(r)^2B(r)}. \label{pt}
\end{align}

The expressions for $\rho$, $p_r$, and $p_t$, given respectively by
Eqs.~(\ref{rho})--(\ref{pt}), were obtained assuming
Eq.~\eqref{Comp_TA} in the region where the temporal coordinate $t$
is timelike, i.e., when $A(r) > 0$.
In the region where $A(r) < 0$, the coordinate $t$ becomes
spacelike, and the mixed components of the anisotropic fluid take
the form
\begin{equation}
	T^{\mu}_{\;\nu} = \text{diag}\left( -p_{r},\, \rho,\, -p_{t},\, -p_{t} \right).
	\label{Comp_TA2}
\end{equation}
From the Einstein equations~\eqref{EqM}, the fluid components
corresponding to Eq.~\eqref{Comp_TA2} are
\begin{align}
	\kappa^2\rho(r) &=
	-\frac{A'(r)\Sigma'(r)}{A(r)B(r)\Sigma(r)}
	- \frac{\Sigma'(r)^2}{B(r)\Sigma(r)^2}
	+ \frac{1}{\Sigma(r)^2}; \label{rhoD} \\[4pt]
	\kappa^2p_r(r) &=
	-\frac{B'(r)\Sigma'(r)}{B(r)^2\Sigma(r)}
	+ \frac{\Sigma'(r)^2}{B(r)\Sigma(r)^2}
	- \frac{1}{\Sigma(r)^2}
	\nonumber \\
	& + \frac{2\Sigma''(r)}{B(r)\Sigma(r)}; \label{prD} \\[4pt]
	\kappa^2p_t(r) &=
	\frac{A'(r)\Sigma'(r)}{2A(r)B(r)\Sigma(r)}
	- \frac{A'(r)B'(r)}{4A(r)B(r)^2}
	+ \frac{\Sigma''(r)}{B(r)\Sigma(r)} \nonumber \\
	&\quad
	+ \frac{A''(r)}{2A(r)B(r)}
	- \frac{B'(r)\Sigma'(r)}{2B(r)^2\Sigma(r)}
	- \frac{A'(r)^2}{4A(r)^2B(r)}. \label{ptD}
\end{align}

Equations~(\ref{rhoD})--(\ref{ptD}) satisfy all the components of
the equations of motion, just as Eqs.~(\ref{rho})--(\ref{pt}) do.

As previously discussed, the matter source tensors given by
Eqs.~\eqref{TNED2} and~\eqref{Tphi2} can be combined to form the
components of a more general anisotropic tensor:
\begin{align}
	\rho(r)   &= {\cal L}(r)
	+ \frac{\epsilon(r)\varphi'(r)^{2}}{B(r)}
	+ V(\varphi); \label{Eq_rho} \\[4pt]
	-p_r(r)   &= {\cal L}(r)
	- \frac{\epsilon(r)\varphi'(r)^{2}}{B(r)}
	+ V(\varphi); \label{Eq_pr0} \\[4pt]
	-p_t(r)   &= {\cal L}(r)
	- \frac{q_m^{2}{\cal L}_{F}(r)}{\Sigma(r)^{4}}
	+ \frac{\epsilon(r)\varphi'(r)^{2}}{B(r)}
	+ V(\varphi). \label{Eq_pt0}
\end{align}
From Eqs.~(\ref{Eq_rho})--(\ref{Eq_pt0}), it is clear that the
matter sources described by the NED and the scalar field can be
interpreted in terms of an effective anisotropic fluid.

Let us now consider the following relation, which will be used to
analyse our solutions under specific restrictions on the matter
sector:
\begin{align}
	\rho(r) + p_r(r) =
	\frac{1}{\kappa^2}\Bigg( 
	\frac{A'(r)\Sigma'(r)}{A(r)B(r)\Sigma(r)}
	+ \frac{B'(r)\Sigma'(r)}{B(r)^{2}\Sigma(r)}
	    &
        \nonumber\\
        - \frac{2\Sigma''(r)}{B(r)\Sigma(r)}
	\Bigg). &\label{rel_pr+rho}
\end{align}
In Eq.~\eqref{rel_pr+rho}, one may impose the restriction
$\rho(r) + p_r(r) = 0$, which typically occurs in static and
spherically symmetric configurations within electromagnetic
theories, whether linear or nonlinear, as can be seen from the
electromagnetic matter sector in Eq.~\eqref{TNED2}, where
$p_r(r) = -\rho(r) = {\cal L}(r)$.  Alternatively, one may
consider a source satisfying $\rho(r) + p_r(r) \neq 0$, without
loss of generality.

In this work, we restrict ourselves to the case
$\rho(r) + p_r(r) = b_1(r) \neq 0$.  For convenience, we rewrite
Eq.~\eqref{rel_pr+rho} as
\begin{equation}
	\kappa^2 b_1(r) =
	\frac{A'(r)\Sigma'(r)}{A(r)B(r)\Sigma(r)}
	+ \frac{B'(r)\Sigma'(r)}{B(r)^{2}\Sigma(r)}
	- \frac{2\Sigma''(r)}{B(r)\Sigma(r)}.
	\label{rel2_pr+rho}
\end{equation}

Integrating Eq.~\eqref{rel2_pr+rho}, we obtain the following
dependence:
\begin{equation}
	B(r) = \frac{\Sigma'(r)^{2}}
	{A(r)\left( 1 - \kappa^2\displaystyle\int
		\frac{b_{1}(r)\Sigma(r)\Sigma'(r)}{A(r)}\,dr
		\right)}.
	\label{B}
\end{equation}
Note that, in the specific case $b_1(r) = 0$, i.e., when
$\rho(r) + p_r(r) = 0$, we recover Eq.~(16) obtained by the authors
in Ref.~\cite{Lessa:2024erf}.

The explicit form of $B(r)$ follows from Eq.~\eqref{B} once
$b_1(r)$, $A(r)$, and the area function $\Sigma(r)$ are specified.
These steps will be demonstrated in the following sections, where we
discuss our models.  Furthermore, once the metric function $A(r)$
is imposed and $B(r)$ is deduced, we can obtain the corresponding
expressions for $\rho(r)$, $p_r(r)$, and $p_t(r)$, allowing us to
analyse the energy conditions for each solution.  Finally, we shall
examine the matter Lagrangian and potential that sustain the
solutions of our models.  To this end, we constrain the form of
$b_1(r)$ in order to ensure consistency with the solutions and to
enable the analytical determination of $B(r)$ from Eq.~\eqref{B}.
This restriction also avoids unphysical situations, such as the
metric exhibiting two timelike components outside the event horizon.

\subsection{Energy conditions}

Finally, we analyse the energy conditions (ECs), which are
fundamental for constraining the matter content and provide a
physical understanding of geodesic focusing.  For the
energy--momentum tensor given in Eq.~\eqref{Comp_TA}, the null
energy condition (NEC), weak energy condition (WEC), strong energy
condition (SEC), and dominant energy condition (DEC) are expressed
by the following inequalities:
\begin{align}
	\mathrm{NEC}_{1,2} &= \rho + p_{r,t} \geq 0, \label{NEC} \\[4pt]
	\mathrm{SEC}_3     &= \rho + p_r + 2p_t \geq 0, \label{SEC} \\[4pt]
	\mathrm{DEC}_{1,2} &= \rho - |p_{r,t}| \geq 0
	\quad \text{or} \quad \rho \pm p_{r,t} \geq 0, \label{DEC} \\[4pt]
	\mathrm{DEC}_3 &= \mathrm{WEC}_3 = \rho \geq 0. \label{WEC}
\end{align}
Here the indices $1$ and $2$ correspond, respectively, to the
radial and tangential components of the anisotropic fluid.  In what
follows, we examine under which parameter conditions the ECs can be
fulfilled in the models presented.

\section{Model}\label{sec:ModeI}

\subsection{Model with $\rho(r)+p_r(r)=b_1(r)$}

In this first model, we consider the following choice for the
function $b_1(r)$:
\begin{equation}
	b_1(r) = \frac{4 b_0^2 \, r A(r)}
	{\kappa^2 \left(q_m^2 + r^2\right)^3
		\Sigma(r)\,\Sigma'(r)} \ ,
	\label{b1}
\end{equation}
where $b_0$ is a constant. Furthermore, we adopt the area function proposed by Simpson and
Visser~\cite{Simpson:2018tsi},
\begin{equation}
	\Sigma(r) = \sqrt{q_m^2 + r^2},
	\label{Sigma}
\end{equation}
together with the following metric function:
\begin{equation}
	A(r) = 1 - \frac{2M}{\Sigma(r)} + \frac{\rho_0}{\Sigma(r)^2}
	= 1 - \frac{2M}{\sqrt{q_m^2 + r^2}}
	+ \frac{\rho_0}{q_m^2 + r^2}.
	\label{AMod1}
\end{equation}
We note that, in the limit $r \to 0$, the metric function depends
only on the constants $M$, $q_m$, and $\rho_0$, while in the limit
$r \to \infty$, it becomes asymptotically Minkowskian.  This metric
function is a special case of Eq.~(24) presented in
Ref.~\cite{Lessa:2024erf} for $\omega = 1$ and $a = q_m$.

Substituting Eqs.~\eqref{b1}, \eqref{Sigma}, and \eqref{AMod1}
into Eq.~\eqref{B}, we obtain the metric function $B(r)$:
\begin{equation}
	B(r)^{-1} =
	\left( 1 - \frac{2M}{\Sigma(r)}
	+ \frac{\rho_0}{\Sigma(r)^2} \right)
	\left( \frac{\Sigma(r)^2}{r^2} \right) \left( 1 + \frac{b_0^2}{\Sigma(r)^4} \right).
	\label{BMod1}
\end{equation}
Note that, if we take $b_0 \to 0$, we recover the case
$\rho(r) + p_r(r) = 0$, which corresponds exactly to the solution
described by Eq.~(25) in Ref.~\cite{Lessa:2024erf} for $\omega = 1$
and $a = q_m$.

\subsubsection{Horizons}

\begin{figure}[htb!]
	\centering
	\includegraphics[scale=0.38]{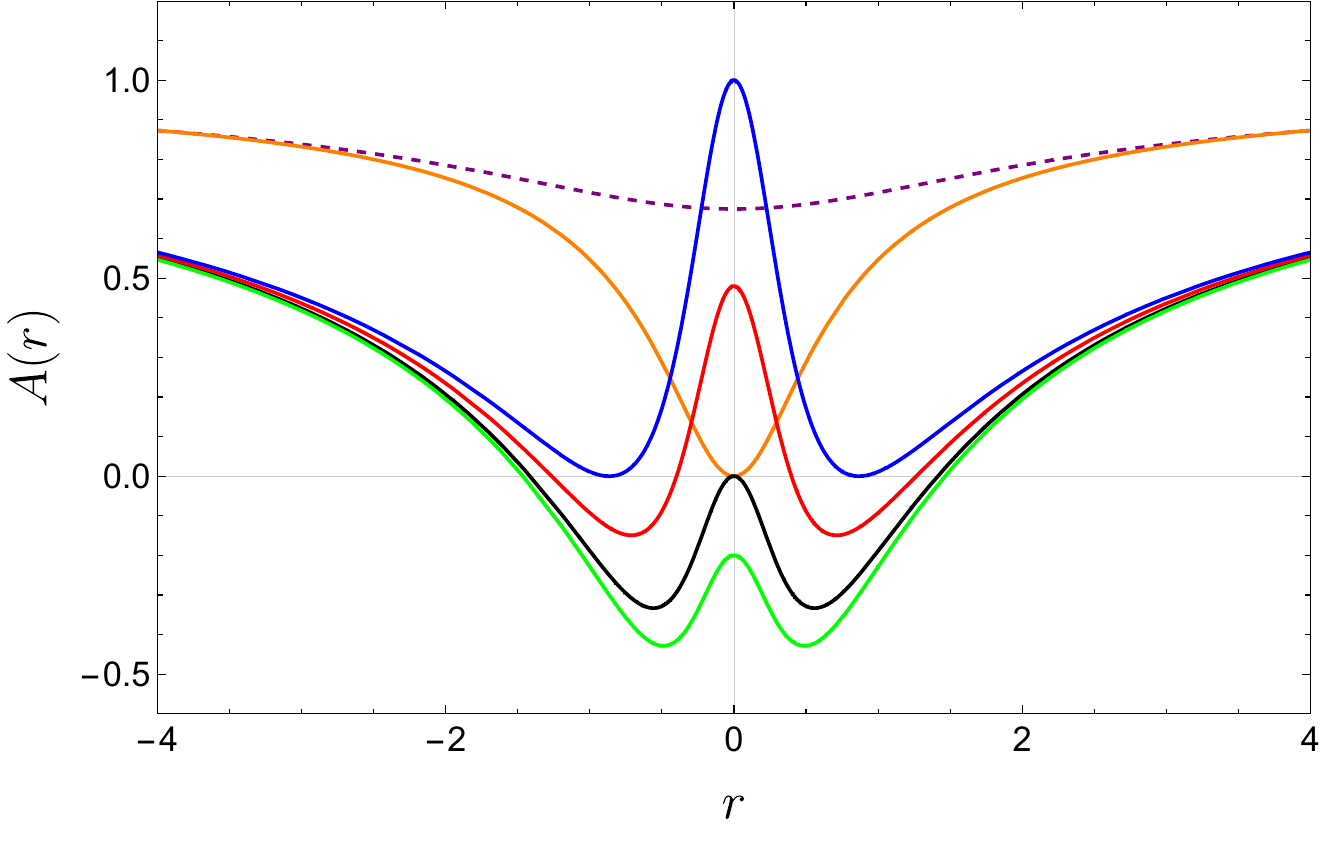} 
	\caption{Behaviour of the metric function given by
		Eq.~\eqref{AMod1}.}
	\label{figA}
\end{figure}

We now examine the intrinsic properties of the metric
function~\eqref{AMod1}.  The presence of horizons is determined
by the condition $A(r_H) = 0$, whose solution is
\begin{equation}
	r_H = \pm\sqrt{\left(M \pm \sqrt{M^2 - \rho_0}\right)^2 - q_m^2}.
	\label{rH}
\end{equation}
The outer pair of $\pm$ signs corresponds, respectively, to the
horizons in our universe ($r > 0$) and in the parallel universe
beyond the throat ($r < 0$).  The inner pair distinguishes the
outer horizon (positive sign) from the inner horizon (negative
sign).  Depending on the values of the parameters, the following
configurations arise:

\begin{itemize}
	\item \textbf{Black bounce:} for $\rho_0 < 2Mq_m - q_m^2$,
	with $q_m < M \pm \sqrt{M^2 - \rho_0}$ and
	$M^2 > \rho_0$, the spacetime possesses two horizons---one
	in the region $r < 0$ and another in $r > 0$---connected
	by a bounce at $r = 0$.  This case is illustrated by the
	solid green curve in Fig.~\ref{figA}, for $\rho_0 = 0.7$,
	$q_m = 0.5$, and $M = 1.0$.
	
	\item \textbf{Black bounce with null throat:} here two
	degenerate horizons are located at the origin, coinciding
	with the throat and forming a null throat configuration.
	This occurs when
	$q_m = \sqrt{M^2 - \rho_0} + M$ and
	$\rho_0 = 2Mq_m - q_m^2$, with $M > \rho_0$.  This case
	is shown by the solid black curve in Fig.~\ref{figA},
	for $\rho_0 = 0.75$, $q_m = 0.5$, and $M = 1.0$.
	
	\item \textbf{Regular black hole with throat:} for
	$M^2 > \rho_0$ and
	$q_m < M \pm \sqrt{M^2 - \rho_0}$, we obtain a regular
	black hole with up to four horizons---the outer ones
	being the event horizons and the inner ones the Cauchy
	horizons.  At $r = 0$, a throat connects the regions
	$r > 0$ and $r < 0$.  This case is represented by the
	solid red curve in Fig.~\ref{figA}, for $q_m = 0.5$,
	$M = 1.0$, and $\rho_0 = 0.85$.
	
	\item \textbf{Wormhole with null throat:} all horizons
	coincide at the origin, $r_H = 0$, forming an extreme
	null throat.  This configuration is depicted by the
	solid orange curve in Fig.~\ref{figA}, for
	$\rho_0 = 0.006$, $q_m = 0.5$, and $M = 0.256$.
	
	\item \textbf{Extreme black bounce:} for $q_m \leq M$, the
	spacetime exhibits degenerate horizons at
	$r_{\rm ext} = \pm\sqrt{M^2 - q_m^2}$, corresponding to
	the critical value $\rho_0 = M^2$, together with a throat
	at $r = 0$.  This case is shown by the solid blue curve
	in Fig.~\ref{figA}, for $\rho_0 = M^2$, $q_m = 0.5$, and
	$M = 1.0$.
	
	\item \textbf{Wormhole:} this spacetime occurs when
	$M^2 > \rho_0$ and
	$q_m > M \pm \sqrt{M^2 - \rho_0}$.  No horizons are
	present, corresponding to a traversable wormhole.  This
	case is illustrated by the dashed purple curve in
	Fig.~\ref{figA}, for $q_m = 1$, $M = 0.25$, and
	$\rho_0 = -0.3$.
\end{itemize}

\subsection{Matter described by an anisotropic fluid and energy conditions}
\label{sec:fluid_EC}

With the metric functions of this model defined by
Eqs.~\eqref{AMod1} and~\eqref{BMod1}, together with the area
function~\eqref{Sigma}, the components of the anisotropic fluid,
$\rho(r)$, $p_r(r)$, and $p_t(r)$, given by
Eqs.~(\ref{rho})--(\ref{pt}), take the  forms:
\begin{align}
	\rho(r) =&
	\frac{b_0^2\left[ -8M\Sigma(r)
		+ 5\rho_0 + 3\Sigma(r)^2 \right]}
	{\kappa^2 \Sigma(r)^8}
		+ \frac{\rho_0}{\kappa^2 \Sigma(r)^4}; \label{rho1} \\[4pt]
	p_r(r) =&
	\frac{b_0^2\left( \Sigma(r)^2 - \rho_0 \right)
		- \rho_0 \Sigma(r)^4}
	{\kappa^2 \Sigma(r)^8}; \label{pr1} \\[4pt]
	p_t(r) =&
	\frac{b_0^2\left[ 2M\Sigma(r)
		+ \rho_0 - 2\Sigma(r)^2 \right]}
	{\kappa^2 \Sigma(r)^8}
	+ \frac{\rho_0}{\kappa^2 \Sigma(r)^4}. \label{pt1}
\end{align}
When $b_0 = 0$, Eqs.~\eqref{rho1}--\eqref{pt1} reduce to the case
$\rho(r) + p_r(r) = 0$.  In the limit $r \to \infty$, all three
components vanish.  Depending on the sign of the constant $\rho_0$,
they may approach zero from negative values, as can be seen from
the asymptotic expansions
\begin{equation}
	\rho(r) \sim \frac{\rho_0}{\kappa^2 r^4}, \qquad
	p_r(r) \sim -\frac{\rho_0}{\kappa^2 r^4}, \qquad
	p_t(r) \sim \frac{\rho_0}{\kappa^2 r^4}.
	\label{asymptotic_rho_p}
\end{equation}

From Eqs.~\eqref{rho1}--\eqref{pt1}, the energy conditions
\eqref{NEC}--\eqref{WEC} outside the event horizon, i.e., in the
region where $A(r) > 0$ and $t$ is the timelike coordinate, are
given by
\begin{align}
	\mathrm{NEC}_1 &= \mathrm{SEC}_1 = \mathrm{WEC}_1 \nonumber \\
	&= \frac{4b_0^2\left( -2M\Sigma(r)
		+ \rho_0 + \Sigma(r)^2 \right)}
	{\kappa^2 \Sigma(r)^8}
	\geq 0; \label{NEC1F_Mod1} \\[4pt]
	\mathrm{NEC}_2 &= \mathrm{SEC}_2 = \mathrm{WEC}_2 
	= \frac{2\rho_0}{\kappa^2 \Sigma(r)^4}
	\nonumber \\
	&
	+ \frac{b_0^2\left( -6M\Sigma(r)
		+ 6\rho_0 + \Sigma(r)^2 \right)}
	{\kappa^2 \Sigma(r)^8}
	\geq 0; \label{NEC2F_Mod1} \\[4pt]
	\mathrm{SEC}_3 &=
	\frac{2b_0^2\left( 3\rho_0 - 2M\Sigma(r) \right)}
	{\kappa^2 \Sigma(r)^8}
	+ \frac{2\rho_0}{\kappa^2 \Sigma(r)^4}
	\geq 0; \label{SEC3F_Mod1} \\[4pt]
	\mathrm{DEC}_1 &=
	\frac{b_0^2\left[ 3\Sigma(r)^2
		- 8M\Sigma(r) + 5\rho_0 \right]}
	{\kappa^2 \Sigma(r)^8}
		+ \frac{\rho_0}{\kappa^2 \Sigma(r)^4} \nonumber \\
	&\quad - \left|
	\frac{b_0^2( -\rho_0 + \Sigma(r)^2 )
		- \rho_0\Sigma(r)^4}
	{\kappa^2 \Sigma(r)^8}
	\right| \geq 0; \label{DEC1F_Mod1} \\[4pt]
	\mathrm{DEC}_2 &=
	\frac{b_0^2\left[ 3\Sigma(r)^2
		- 8M\Sigma(r) + 5\rho_0 \right]}
	{\kappa^2 \Sigma(r)^8}
	+ \frac{\rho_0}{\kappa^2 \Sigma(r)^4} \nonumber \\
	& - \left|
	\frac{b_0^2\left[ 2M\Sigma(r)
		+ \rho_0 - 2\Sigma(r)^2 \right]}
	{\kappa^2 \Sigma(r)^8}+ \frac{\rho_0}{\kappa^2 \Sigma(r)^4}
	\right| \geq 0; \label{DEC2F_Mod1} \\[4pt]
	\mathrm{DEC}_3 &= \mathrm{WEC}_3 
	= \frac{\rho_0}{\kappa^2 \Sigma(r)^4}
	+\nonumber \\
	& \frac{b_0^2\left[ -8M\Sigma(r)
		+ 5\rho_0 + 3\Sigma(r)^2 \right]}
	{\kappa^2 \Sigma(r)^8}
	\geq 0. \label{DEC3F_Mod1}
\end{align}
We note that, in the limit $r \to \infty$, all of these energy conditions tend to zero, so that for a distant observer they are
always satisfied.

To perform a more detailed analysis of their behaviour, we examine
the asymptotic expansions for large $r$.  For $r \gg 1$, we obtain
\begin{align}
	\mathrm{NEC}_1 &\sim \frac{4b_0^2}{\kappa^2 r^6}
	+ \mathcal{O}\!\left( \frac{1}{r^7} \right), \\[4pt]
	\mathrm{NEC}_2 &\sim \frac{2\rho_0}{\kappa^2 r^4}
	+ \mathcal{O}\!\left( \frac{1}{r^5} \right), \\[4pt]
	\mathrm{SEC}_3 &\sim \frac{2\rho_0}{\kappa^2 r^4}
	+ \mathcal{O}\!\left( \frac{1}{r^5} \right), \\[4pt]
	\mathrm{DEC}_1 &\sim \frac{4b_0^2}{\kappa^2 r^6}
	+ \mathcal{O}\!\left( \frac{1}{r^7} \right), \\[4pt]
	\mathrm{DEC}_2 &\sim \frac{5b_0^2}{\kappa^2 r^6}
	+ \mathcal{O}\!\left( \frac{1}{r^6} \right), \\[4pt]
	\mathrm{DEC}_3 &\sim \frac{\rho_0}{\kappa^2 r^4}
	+ \mathcal{O}\!\left( \frac{1}{r^5} \right).
\end{align}
For sufficiently large $r$, the dominant terms in these expansions
are negative only if $\rho_0 < 0$; otherwise, for $\rho_0 > 0$,
the energy conditions remain positive.  In the specific cases of
$\mathrm{DEC}_1$ and $\mathrm{DEC}_2$, the presence of the $b_0$
term guarantees that these conditions approach zero from positive
values, regardless of the sign of $b_0$.  Hence, for all energy
conditions to approach zero from positive values as
$r \to \pm\infty$, it is sufficient that $\rho_0 > 0$.  In
summary, all the energy conditions are automatically satisfied in
the asymptotic regime $r \to \pm\infty$, whenever $A(r) > 0$,
provided that $\rho_0 > 0$, and this conclusion holds irrespective
of the value of $b_0$.

From Eqs.~\eqref{rhoD}--\eqref{ptD}, the fluid components inside
the event horizon, i.e., where $A(r) < 0$, take the following forms
for this model:
\begin{align}
	\rho(r) &=
	-\frac{b_0^2\left( \Sigma(r)^2 - \rho_0 \right)
		- \rho_0\Sigma(r)^4}
	{\kappa^2\Sigma(r)^8}; \label{rho1D} \\[4pt]
	p_r(r) &=
	-\frac{b_0^2\left[ -8M\Sigma(r)
		+ 5\rho_0 + 3\Sigma(r)^2 \right]}
	{\kappa^2\Sigma(r)^8}\nonumber \\
	&- \frac{\rho_0}{\kappa^2\Sigma(r)^4}; \label{pr1D} \\[4pt]
	p_t(r) &=
	\frac{b_0^2\left[ 2M \Sigma(r)
		+ \rho_0 - 2\Sigma(r)^2 \right]}
	{\kappa^2\Sigma(r)^8}\nonumber \\
	&+ \frac{\rho_0}{\kappa^2\Sigma(r)^4}. \label{pt1D}
\end{align}

In the limit $r \to 0$, these components depend only on the
constants of the model:
\begin{align}
	\lim_{r \to 0} \rho(r) &=
	\frac{b_0^2\left( \rho_0 - q_m^2 \right)
		+ \rho_0 q_m^4}{\kappa^2 q_m^8}, \\[4pt]
	\lim_{r \to 0} p_r(r) &=
	\frac{b_0^2\left( 8M q_m - 5\rho_0 - 3q_m^2 \right)
		- \rho_0 q_m^4}{\kappa^2 q_m^8}, \\[4pt]
	\lim_{r \to 0} p_t(r) &=
	\frac{b_0^2\left( 2M q_m + \rho_0 - 2q_m^2 \right)
		+ \rho_0 q_m^4}{\kappa^2 q_m^8}.
\end{align}

The energy conditions inside the event horizon, i.e., when
$A(r) < 0$, are given by
\begin{align}
	\mathrm{NEC}_1 &= \mathrm{SEC}_1 = \mathrm{WEC}_1 \nonumber \\
	&= -\frac{4b_0^2\left( -2M\Sigma(r)
		+ \rho_0 + \Sigma(r)^2 \right)}
	{\kappa^2\Sigma(r)^8}
	\geq 0; \label{NEC1D_Mod1} \\[4pt]
	\mathrm{NEC}_2 &= \mathrm{SEC}_2 = \mathrm{WEC}_2  \label{NEC2D_Mod1}\\
	&= \frac{b_0^2\left[ 2M\Sigma(r)
		+ 2\rho_0 - 3\Sigma(r)^2 \right]}
	{\kappa^2\Sigma(r)^8}
	+ \frac{2\rho_0}{\kappa^2\Sigma(r)^4}
	\geq 0; \nonumber  \\[4pt]
	\mathrm{SEC}_3 &=
	\frac{2b_0^2}{\kappa^2} \left[
	\frac{6M}{\Sigma(r)^{7}}
	- \frac{\rho_0 + 4\Sigma(r)^2}
	{\Sigma(r)^8}
	\right]
	+ \frac{2\rho_0}{\kappa^2\Sigma(r)^4}
	\geq 0; \label{SEC3D_Mod1} 
    \end{align}
    \begin{align}
	\mathrm{DEC}_1 & =
	\frac{\rho_0\Sigma(r)^4
		- b_0^2\left( \Sigma(r)^2- \rho_0 \right)}
	{\kappa^2\Sigma(r)^8}  \label{DEC1D_Mod1}\\
	& - \Bigg|
	\frac{b_0^2\left[ 5\rho_0
		- 8M\Sigma(r)
		+ 3\Sigma(r)^2 \right]}
	{\kappa^2\Sigma(r)^8}
	+ \frac{\rho_0}{\kappa^2\Sigma(r)^4}
	\Bigg| \geq 0; \nonumber \\[4pt]
	\mathrm{DEC}_2 &=
	\frac{\rho_0\Sigma(r)^4
		- b_0^2\left( \Sigma(r)^2 - \rho_0 \right)}
	{\kappa^2\Sigma(r)^8}  \label{DEC2D_Mod1} \\
	&\quad - \Bigg|
	\frac{b_0^2\left[ 2M\Sigma(r)
		+ \rho_0 - 2\Sigma(r)^2 \right]}
	{\kappa^2\Sigma(r)^8}
	+ \frac{\rho_0}{\kappa^2\Sigma(r)^4}
	\Bigg| \geq 0; \nonumber \\[4pt]
	\mathrm{DEC}_3 &= \mathrm{WEC}_3 = -\frac{b_0^2\left( \Sigma(r)^2 - \rho_0 \right)
		- \rho_0\Sigma(r)^4}
	{\kappa^2\Sigma(r)^8}
	\geq 0. \label{DEC3D_Mod1}
\end{align}

To examine the energy conditions inside the event horizon, we
assume that the constants $q_m$, $M$, and $b_0$ take positive,
well-defined values.  We then determine the constraints that the
parameter $\rho_0$ must satisfy in order for all energy conditions
to hold.

Considering very small values of $r$ in the expression for
$\mathrm{NEC}_1$ inside the event horizon, given by
Eq.~\eqref{NEC1D_Mod1}, we find
\begin{equation}
	\lim_{r \to 0} \mathrm{NEC}_1
	= \frac{4b_0^2}{\kappa^2} \left(
	\frac{2M}{q_m^7} - \frac{\rho_0 + q_m^2}{q_m^8}
	\right).
\end{equation}
From this limit, we deduce the constraint
$\rho_0 \leq 2Mq_m - q_m^2$.  Thus, for fixed $q_m$ and $M$, this
inequality determines the values of $\rho_0$ for which
$\mathrm{NEC}_1$ is satisfied at the centre.  If $2M > q_m$, then
$\rho_0 > 0$; if $2M < q_m$, then $\rho_0 < 0$; and in the
particular case $q_m = 2M$, $\rho_0$ vanishes and $\mathrm{NEC}_1$
is satisfied at the centre.  We note that, by setting $\rho_0 = 0$
directly in Eq.~\eqref{NEC1D_Mod1}, the condition
$\mathrm{NEC}_1 \geq 0$ for small $r$ reduces to
$M \geq q_m/2$.

For $\mathrm{NEC}_2$ inside the event horizon,
Eq.~\eqref{NEC2D_Mod1}, the limit $r \to 0$ yields
\begin{equation}
	\lim_{r \to 0} \mathrm{NEC}_2 =
	\frac{b_0^2\left[ 2\left( M q_m + \rho_0 \right)
		- 3q_m^2 \right] + 2\rho_0 q_m^4}
	{\kappa^2 q_m^8}.
\end{equation}
This energy condition is satisfied at the centre if
$\rho_0 \geq \dfrac{b_0^2 q_m(3q_m - 2M)}{2\left( b_0^2 + q_m^4 \right)}$,
for fixed positive values of $q_m$, $M$, and $b_0$.  It follows
that if $q_m > 2M/3$, then $\rho_0 > 0$; if $q_m < 2M/3$, then
$\rho_0 < 0$; and in the case $q_m = 2M/3$, $\rho_0 = 0$.
Setting $\rho_0 = 0$, $\mathrm{NEC}_2$ is violated for small $r$
if and only if $q_m > 2M/3$.

For $\mathrm{SEC}_3$ inside the event horizon,
Eq.~\eqref{SEC3D_Mod1}, the limit $r \to 0$ gives
\begin{equation}
	\lim_{r \to 0} \mathrm{SEC}_3 =
	\frac{2\rho_0 q_m^4
		- 2b_0^2\left( \rho_0 - 6M q_m + 4q_m^2 \right)}
	{\kappa^2 q_m^8}.
\end{equation}
The corresponding constraint for this energy condition to be
satisfied is
$\rho_0 \leq \dfrac{2b_0^2 q_m(3M - 2q_m)}{b_0^2 - q_m^4}$.
When $M < 2q_m/3$, the numerator is negative; in this case,
$\rho_0 < 0$ if $b_0^2 > q_m^4$, whereas $\rho_0 > 0$ if
$b_0^2 < q_m^4$.  Conversely, if $M > 2q_m/3$, the numerator is
positive, and $\rho_0 > 0$ for $b_0^2 > q_m^4$ and $\rho_0 < 0$
for $b_0^2 < q_m^4$.  In the limiting case $M = 2q_m/3$, the
numerator vanishes and the constraint reduces to
$\rho_0\left( q_m^4 - b_0^2 \right) \geq 0$.  Thus, $\rho_0 > 0$
if $q_m^4 > b_0^2$, $\rho_0 < 0$ if $q_m^4 < b_0^2$, and
$\rho_0 = 0$ only for $b_0^2 = q_m^4$, leading to a vanishing
$\mathrm{SEC}_3$ at the centre.  For fixed constants, any value
of $\rho_0$ below the established limit results in a violation
of $\mathrm{SEC}_3$ inside the event horizon.  Setting
$\rho_0 = 0$ directly in Eq.~\eqref{SEC3D_Mod1}, the condition
$\mathrm{SEC}_3 \geq 0$ near the centre requires $3M > 2q_m$;
otherwise, $\mathrm{SEC}_3$ is violated within the horizon.

For $\mathrm{DEC}_1$, the limit $r \to 0$ in
Eq.~\eqref{DEC1D_Mod1} yields
\begin{equation}
	\lim_{r \to 0} \mathrm{DEC}_1 =
	-\frac{4b_0^2\left( \rho_0 - 2M q_m + q_m^2 \right)}
	{\kappa^2 q_m^8}.
\end{equation}
From this result, we obtain the constraint
$\rho_0 \leq 2M q_m - q_m^2$, which is positive whenever
$M > q_m/2$.  Setting $\rho_0 = 0$,
Eq.~\eqref{DEC1D_Mod1} indicates that this energy condition is
violated for small $r$ if and only if $q_m > 2M$.

For $\mathrm{DEC}_2$, the limit $r \to 0$ in
Eq.~\eqref{DEC2D_Mod1} gives
\begin{equation}
	\lim_{r \to 0} \mathrm{DEC}_2 =
	\frac{b_0^2}{\kappa^2 q_m^6} \left( 1 - \frac{2M}{q_m} \right),
\end{equation}
which is independent of $\rho_0$.  This condition is satisfied at
the centre whenever $q_m > 2M$, and violated for $q_m < 2M$.
Setting $\rho_0 = 0$, the condition $\mathrm{DEC}_2 \geq 0$ for
small $r$ requires $128 M^6 (q_m - M) < q_m^7$.

Finally, for $\mathrm{DEC}_3$, the limit $r \to 0$ in
Eq.~\eqref{DEC3D_Mod1} yields
\begin{equation}
	\lim_{r \to 0} \mathrm{DEC}_3 =
	\frac{b_0^2\left( \rho_0 - q_m^2 \right)
		+ \rho_0 q_m^4}{\kappa^2 q_m^8}.
\end{equation}
The condition $\mathrm{DEC}_3 \geq 0$ at the centre therefore
requires
$\rho_0 \geq \dfrac{b_0^2 q_m^2}{b_0^2 + q_m^4}$,
which implies $\rho_0 > 0$.  Any value of $\rho_0$ above this
bound guarantees that $\mathrm{DEC}_3$ remains positive in the
central region.  Setting $\rho_0 = 0$ directly in
Eq.~\eqref{DEC3F_Mod1}, the limit $r \to 0$ becomes
$-\dfrac{b_0^2}{\kappa^2 q_m^6}$, which is intrinsically negative,
so that $\mathrm{DEC}_3$ is always violated at the centre in the
absence of the $\rho_0$ term.

\begin{figure*}[htb!]
	\centering
	\subfigure[Energy conditions for a regular black hole.]{
		\includegraphics[scale=0.55]{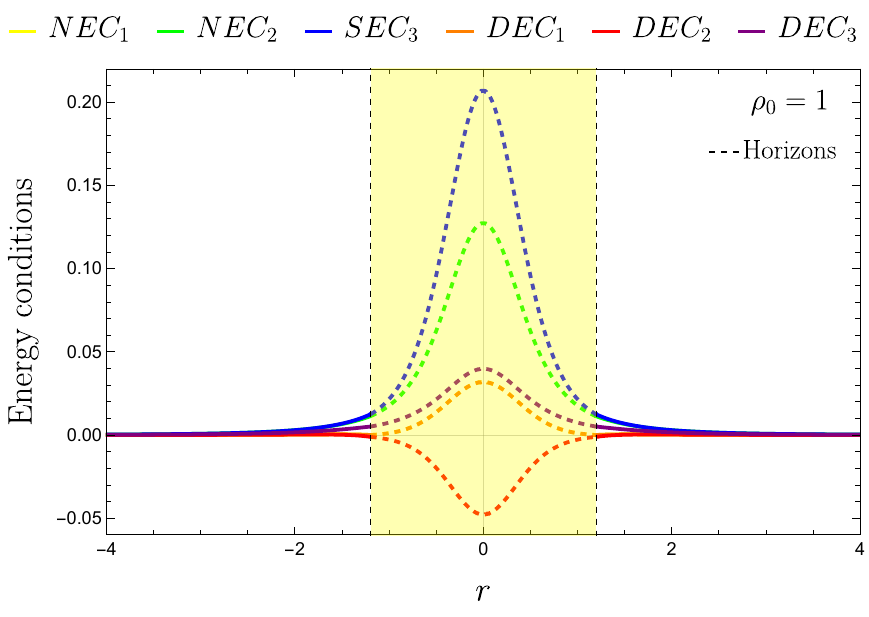}
		\label{fig:EC}
	}
	\hspace{0.75cm}
	\vspace{0.5cm}
	\subfigure[Energy conditions for a regular black hole.]{
		\includegraphics[scale=0.55]{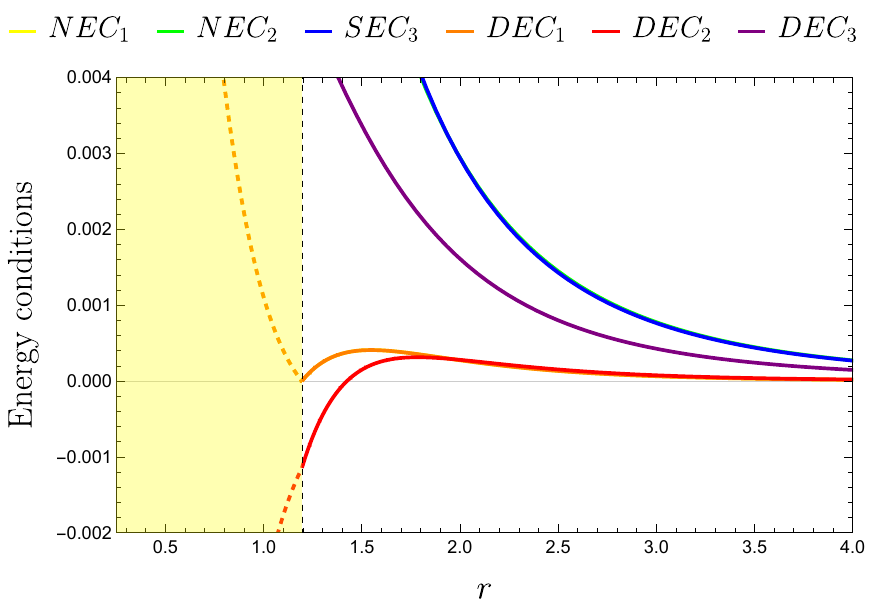}
		\label{fig:EC2}
	}
	\subfigure[Energy conditions for a wormhole.]{
		\includegraphics[scale=0.5]{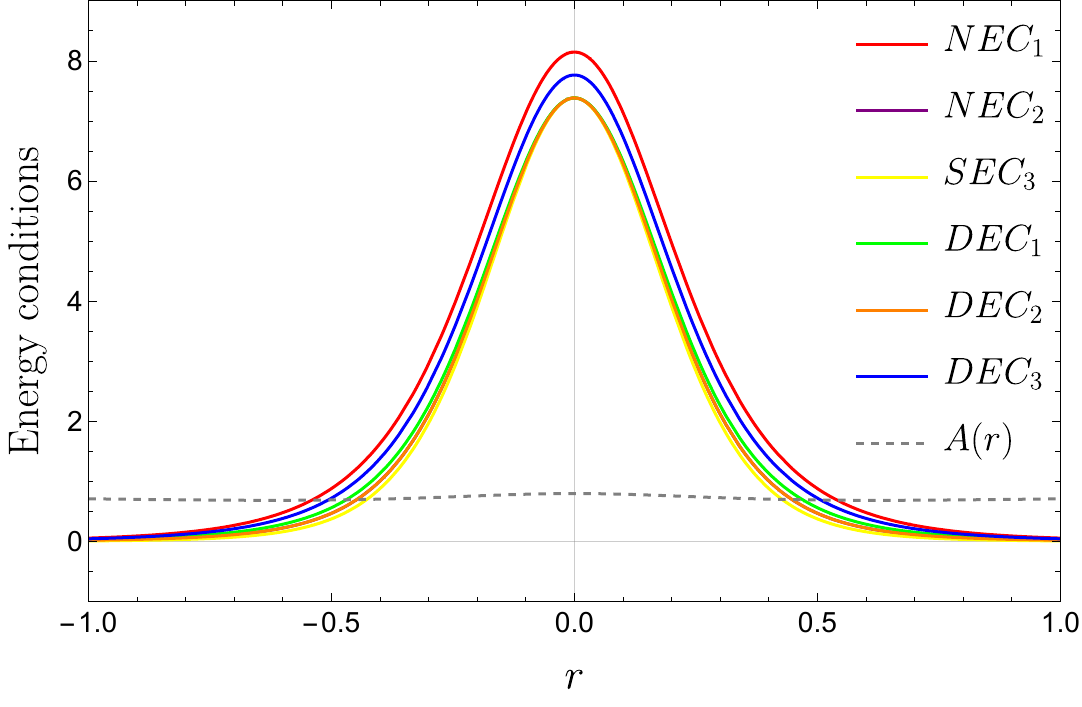}
		\label{fig:worm}
	}
	\caption{Behaviour of the energy conditions.}
	\label{FigEC}
\end{figure*}

The behaviour of the energy conditions outside the event horizon,
given by Eqs.~\eqref{NEC1F_Mod1}--\eqref{DEC3F_Mod1}, and inside
the horizon, described by
Eqs.~\eqref{NEC1D_Mod1}--\eqref{DEC3D_Mod1}, is illustrated in
Figs.~\ref{fig:EC} and~\ref{fig:EC2}.  For these plots, we adopt
the parameter set $b_0 = 1$, $M = 1.1\,q_m$, $q_m = 1$,
$\kappa = \sqrt{8\pi}$, and $\rho_0 = 1$.  For these values, the
metric function~\eqref{AMod1} describes a regular black hole
configuration, with the event horizon located at approximately
$r_H \sim 1.2$.  The black dashed vertical lines mark the
positions of the horizons (in $r < 0$ and $r > 0$), while the
yellow-shaded region indicates the interior of the horizon.  The
solid coloured curves correspond to the energy conditions outside
the event horizon, Eqs.~\eqref{NEC1F_Mod1}--\eqref{DEC3F_Mod1},
whereas the coloured dashed curves within the yellow region
represent the energy conditions inside $r_H$,
Eqs.~\eqref{NEC1D_Mod1}--\eqref{DEC3D_Mod1}.  In
Fig.~\ref{fig:EC2}, $\mathrm{DEC}_2$ (solid red curve) is
slightly violated very close to $r_H$, but as the radial
coordinate increases it approaches zero from positive values,
indicating that the condition becomes satisfied asymptotically.

We also illustrate the energy conditions for the wormhole
configuration, in which no event horizon is present and only
Eqs.~\eqref{NEC1F_Mod1}--\eqref{DEC3F_Mod1} are required.  This
scenario is obtained by setting $b_0 = 1$, $M = q_m/2$,
$q_m = 0.5$, $\kappa = \sqrt{8\pi}$, and $\rho_0 = 0.2$.  As
shown in Fig.~\ref{fig:worm}, the solid curves represent the
energy conditions given by
Eqs.~\eqref{NEC1F_Mod1}--\eqref{DEC3F_Mod1}, while the grey
dashed curve depicts the metric function~\eqref{AMod1}.  We note
that in both configurations shown in Fig.~\ref{FigEC}, the
condition $\mathrm{SEC}_3$ is always satisfied, which is
remarkable since this condition is usually violated in such
scenarios.

\subsection{Regularity check: Kretschmann scalar}

To verify the regularity of a spacetime, it is customary to employ the Kretschmann scalar, defined as $K = R_{\rho\sigma\mu\nu}R^{\rho\sigma\mu\nu}$, and look at its finiteness. For the metric~\eqref{m}, its explicit general form is given by
\begin{align}
	&K(r)=\bigg\{ B(r)^{2}\Sigma(r)^{4}A'(r)^{4}+2A(r)B(r)\Sigma(r)^{4}A'(r)^{2}
	\nonumber\\&    \times
	\Big(A'(r)B'(r)-2B(r)A''(r)\Big)
	+A(r)^{2}\bigg[8B(r)^{2}\Sigma(r)^{2}
	\nonumber
	\\    &
	\times A'(r)^{2}\Sigma'(r)^{2}    +\Sigma(r)^{4}\Big(A'(r)B'(r)-2B(r)A''(r)\Big)^{2}\bigg]
	\nonumber
	\\&  
	+8A(r)^{4}\bigg[2B(r)^{4}
	-4B(r)^{3}\Sigma'(r)^{2}+\Sigma(r)^{2}B'(r)^{2}\Sigma'(r)^{2}
	\nonumber\\&    +2B(r)^{2}\left(2\Sigma(r)^{2}\Sigma''(r)^{2}+\Sigma'(r)^{4}\right)-4B(r)\Sigma(r)^{2}B'(r)
	\nonumber
	\\&    \times\Sigma'(r)\Sigma''(r)\bigg]\bigg\}\bigg/\bigg(4A(r)^{4}B(r)^{4}\Sigma(r)^{4}\bigg)\,.\label{K}
\end{align}
Together with expressions~\eqref{L_BB}, \eqref{LF_BB}, \eqref{eps}, \eqref{V}, \eqref{W}, and~\eqref{L}, Eq.~\eqref{K} completes the set of relevant quantities needed to model a solution with a purely magnetic source. \\
Now, to test the global regularity of the specific model considered above, we examine the behavior of $K(r)$ at the boundaries. Taking the limit $r\to 0$ of Eq.~\eqref{K}, we find
 \begin{align}
    &\lim_{r\to 0} K =\frac{1}{q^{16}}\bigg\{8q_{m}^{8}\left(6M^{2}q_{m}^{2}-12M\rho_{0}\sqrt{q_{m}^{2}}+7\rho_{0}^{2}\right)
    \nonumber
     \\
    &    +8b_{0}^{2}q_{m}^{4}\Big[24M^{2}q_{m}^{2}-2M\sqrt{q_{m}^{2}}\left(23\rho_{0}+3q_{m}^{2}\right)    
    \nonumber
     \\
    & +\rho_{0}\left(24\rho_{0}+5q_{m}^{2}\right)\Big]+4b_{0}^{4}\Big[-4M\sqrt{q_{m}^{2}}\left(27\rho_{0}+11q_{m}^{2}\right)
     \nonumber
     \\
     &     +72M^{2}q_{m}^{2}+46\rho_{0}^{2}+9q_{m}^{4}+26\rho_{0}q_{m}^{2}\Big]\bigg\}
     \,.
 \end{align}
From this expression, we see that in the core region ($r\to 0$), the Kretschmann scalar depends exclusively on the constant parameters of the model and is manifestly finite. Furthermore, in the asymptotic limit $r\to\infty$, the scalar vanishes, as expected for an asymptotically flat spacetime. Taken together, these limiting behaviors confirm that the proposed model remains regular throughout the entire spacetime manifold.

\subsection{Reconstruction of the matter Lagrangian and potential}

We now present the matter content described by the LED and the
scalar field that supports the model under consideration, as
introduced at the beginning of Sec.~\ref{sec:ModeI}.
Recall that we have adopted the choice ${\cal L}_F = 1$, as
stated in Sec.~\ref{sec2}, which implies that we are dealing
with the linear electrodynamics (LED) case.  Substituting the
area function~\eqref{Sigma} into the Lagrangian
density~\eqref{L_BB}, we obtain
\begin{equation}
	{\cal L}(r) = \frac{q_m^2}{2\left( q_m^2 + r^2 \right)^2},
	\label{L1}
\end{equation}
or, equivalently,
\begin{equation}
	{\cal L}(F) = F.
	\label{L2}
\end{equation}

\begin{figure}[htb!]
	\centering
	\includegraphics[scale=0.465]{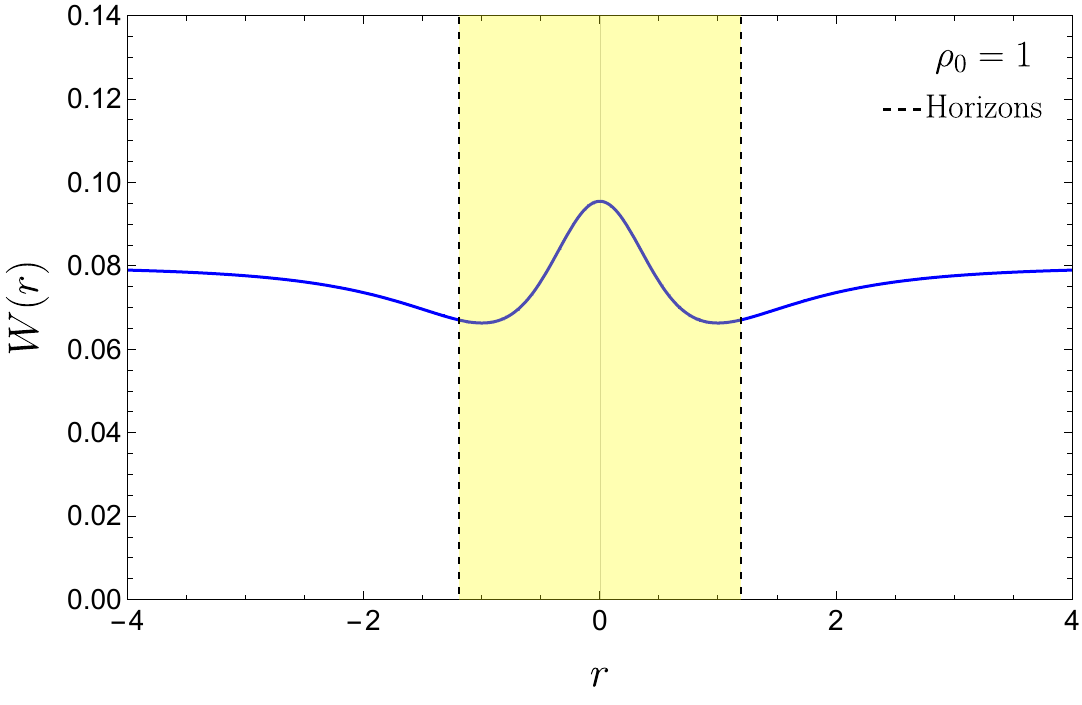} 
	\caption{Behaviour of the coupling function $W(r)$, given by
		Eq.~\eqref{W1}.}
	\label{figw}
\end{figure}

Substituting the metric functions~\eqref{AMod1}
and~\eqref{BMod1}, together with the area
function~\eqref{Sigma}, into Eq.~\eqref{W}, we obtain the
following expression for the coupling function $W(r)$:
\begin{equation}
    W(r)=\frac{2\rho_{0}}{\kappa^{2}q_{m}^{2}}+\frac{b_{0}^{2}\left(6\rho_{0}-6M\Sigma(r)+\Sigma(r)^{2}\right)}{\kappa^{2}q_{m}^{2}\Sigma(r)^{4}}.\label{W1}
\end{equation}
The behaviour of $W(r)$ is illustrated in Fig.~\ref{figw} for
the parameter set $b_0 = 1$, $M = 1.1\,q_m$, $q_m = 1$,
$\kappa = \sqrt{8\pi}$, and $\rho_0 = 1$.  We note that
$W(r)$ remains positive throughout the entire range of the
radial coordinate.

The product $W(r){\cal L}(r)$, which defines the interaction
term in the action, takes the explicit form
\begin{align}
	W(r){\cal L}(r) =&\frac{b_{0}^{2}\left(6\rho_{0}-6M\Sigma(r)+\Sigma(r)^{2}\right)}{2\kappa^{2}\Sigma(r)^{8}}
		+ \frac{2\rho_{0}}{\kappa^{2}\Sigma(r)^{2}}.
\end{align}

\begin{figure}[htb!]
\centering
\includegraphics[scale=0.465]{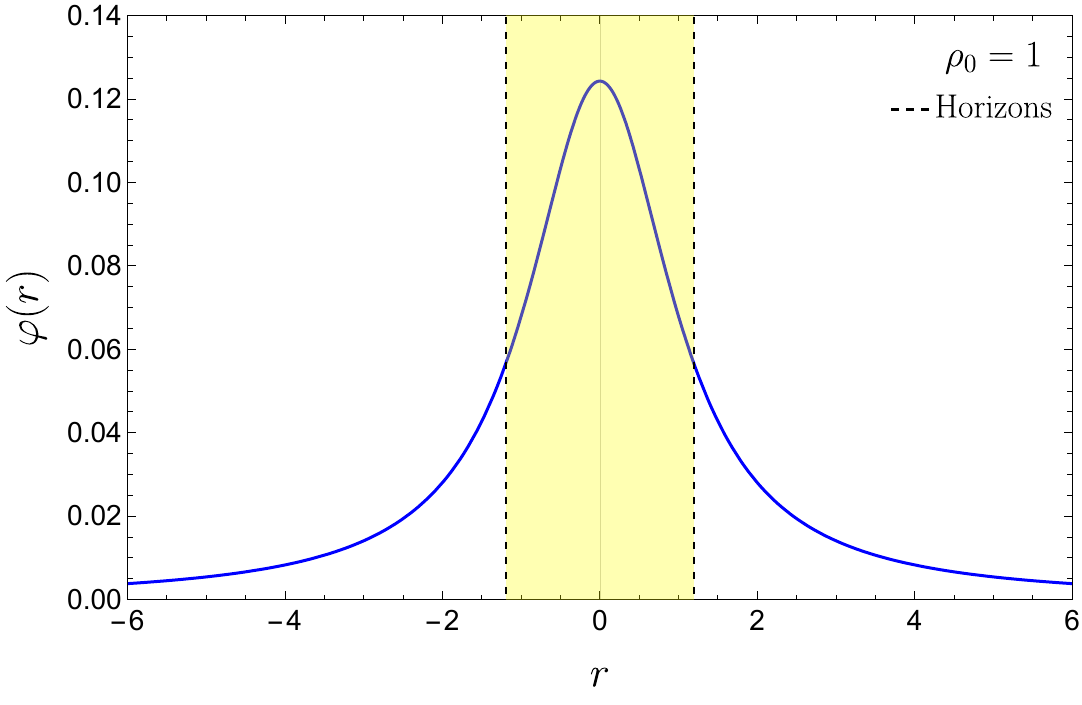} 
\caption{Scalar field behavior, as described by Eq.~\eqref{phiMod1}. }
\label{figphi}
\end{figure}

Once the metric functions $A(r)$ and $\Sigma(r)$ have been
defined---respectively by Eqs.~\eqref{AMod1}
and~\eqref{Sigma}---and $B(r)$ has been determined from
Eq.~\eqref{BMod1}, we can proceed to obtain the scalar field
$\varphi(r)$ from Eq.~\eqref{eps}.  To do so, we must first
specify the form of $\epsilon(r)$.
It is well known that standard black bounce solutions,
characterised by the symmetry $g_{00} = -g_{11}^{-1}$, are
supported by a \textit{phantom scalar field}, corresponding to
$\epsilon(r) = -1$.  In the present model, however,
$g_{00} \neq -g_{11}^{-1}$, which allows us to adopt a
canonical scalar field, i.e., $\epsilon(r) = 1$.
Substituting Eqs.~\eqref{AMod1}, \eqref{BMod1},
and~\eqref{Sigma} into Eq.~\eqref{eps} with $\epsilon(r) = 1$
yields
\begin{equation}
    1=\frac{2b_{0}^{2}r^{2}}{\kappa^{2}\Sigma(r)^{4}\left(b_{0}^{2}+\Sigma(r)^{4}\right)\varphi'(r)^{2}}.
    \label{eps2}
\end{equation}
Solving for $\varphi(r)$, we obtain 
\begin{eqnarray}
	\varphi(r) &=&\frac{1}{2\sqrt{2}\,\kappa}\ln\left[\frac{\sqrt{b_{0}^{2}+\Sigma(r)^{4}}+b_{0}}{\sqrt{b_{0}^{2}+\Sigma(r)^{4}}-b_{0}}\right]
    \nonumber \\
	&=&\frac{1}{\sqrt{2}\,\kappa}\operatorname{arcsinh}\left[\frac{b_0}{\Sigma(r)^{2}}\right]\,.
	\label{phiMod1}
\end{eqnarray}
In the limit $r \to 0$, Eq.~\eqref{phiMod1} reduces to
\begin{align}
	\varphi_0=\frac{1}{\sqrt{2}\,\kappa}\operatorname{arcsinh}\left[\frac{b_0}{q_{m}^{2}}\right] \,,
\end{align}
whereas for $r \to \infty$, the scalar field tends to zero.  We thus see that $b_0$ can be written as $b_0=q_{m}^{2} \sinh[\sqrt{2}\,\kappa \varphi_0]$, with $\varphi_0$ representing the value of the scalar field at the throat. 
The behaviour of $\varphi(r)$ is illustrated in Fig.~\ref{figphi}
for $b_0 = 1$, $M = 1.1\,q_m$, $q_m = 1$,
$\kappa = \sqrt{8\pi}$, and $\rho_0 = 1$ and represents a lump of energy.


With the scalar field $\varphi(r)$ given by
Eq.~\eqref{phiMod1} and the coupling function $W(r)$ given by
Eq.~\eqref{W1}, we can now determine the scalar potential
$V(r)$ from Eq.~\eqref{V}.  For the present model, this yields
\begin{equation}
	V(r)=\frac{b_{0}^{2}\left(\Sigma(r)-2M\right)}{2\kappa^{2}\Sigma(r)^{7}}.
	\label{VMod1}
\end{equation}

An interesting feature arises when we take the limit $b_0 \to 0$
in Eq.~\eqref{eps2}, which corresponds to the condition
$\rho(r) + p_r(r) = 0$.  In this limit, it is not possible to
determine a scalar potential $V(r)$ from the field equations,
and the scalar field itself becomes undetermined---as can be
seen directly from Eq.~\eqref{VMod1}, which vanishes for
$b_0 \to 0$.  This is consistent with the fact that the
condition $\rho(r) + p_r(r) = 0$ reduces the general
expression~\eqref{B} to $B(r) = \Sigma'(r)^2/A(r)$, which, as
shown in Ref.~\cite{Lessa:2024erf}, corresponds to a black
bounce solution supported solely by a nonlinear
electrodynamics source.  We emphasise, however, that this does
not preclude the introduction of a scalar field directly in the
action~\eqref{action} via the kinetic term.

Using the scalar field profile~\eqref{phiMod1}, we can invert
the relation to express $r$ in terms of $\varphi$ and thereby
write the coupling function and the potential as explicit
functions of the scalar field.  The resulting expression for
the coupling function is 
\begin{align}
W(\varphi)= &  \frac{2 \rho_0}{\kappa ^2 q_m^2} +\frac{b_0 \sinh \left(\sqrt{2} \kappa  \varphi \right)}{\kappa ^2 q_m^2} \nonumber\\
	& -\frac{6 b_0 M \sinh ^{\frac{3}{2}}\left(\sqrt{2} \kappa  \varphi \right)-6 \rho_0 \sinh ^2\left(\sqrt{2} \kappa  \varphi \right)}{\kappa ^2 q_m^2}  \ .\label{W2}
\end{align}

\begin{figure}[htb!]
	\centering
	\includegraphics[scale=0.465]{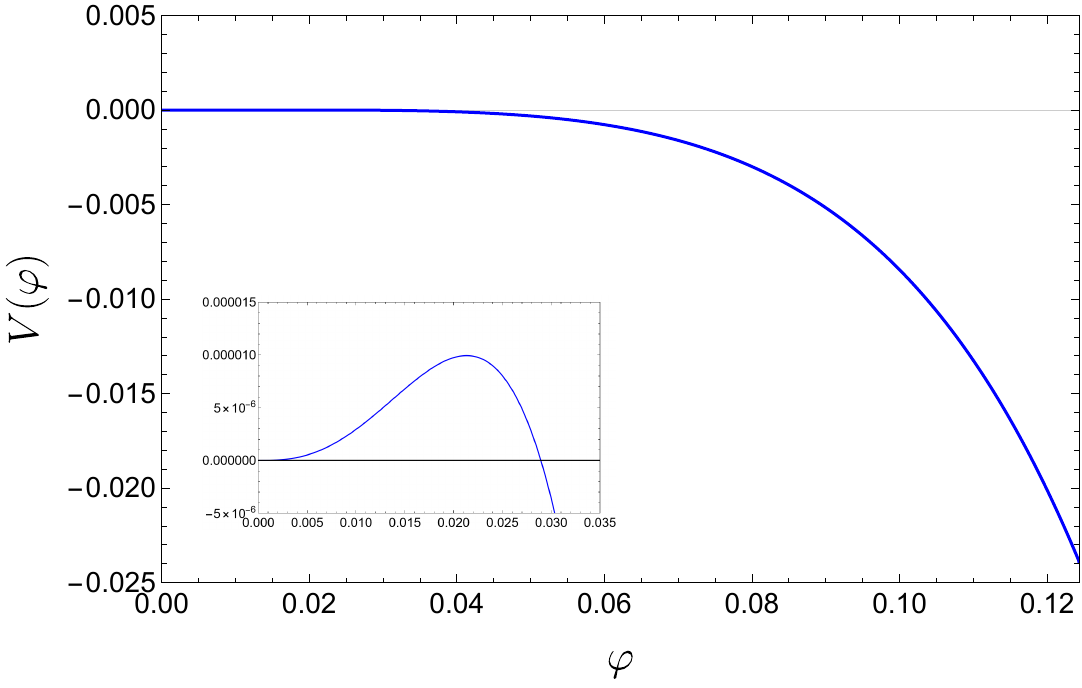} 
	\caption{Behaviour of the scalar potential $V(\varphi)$, given by
		Eq.~\eqref{V2Mod1}.}
	\label{Vq1}
\end{figure}

Similarly, the scalar potential expressed in terms of
$\varphi$ reads
\begin{equation}
	V(\varphi)=   \frac{\sqrt[4]{b_0} \sinh ^{\frac{5}{4}}\left(\sqrt{2} \kappa  \varphi \right) \left(\sqrt{b_0}-2 M {\sinh^{\frac{1}{2}} \left(\sqrt{2} \kappa  \varphi \right)}\right)}{2 \kappa ^2}
	 \ . \label{V2Mod1}
\end{equation}
The behaviour of this potential is illustrated in
Fig.~\ref{Vq1} for $b_0 = 1$, $M = 1.1\,q_m$, $q_m = 1$,
$\kappa = \sqrt{8\pi}$, and $\rho_0 = 1$.

\section{Thin-shell matching and resolution of the energy condition violations}
\label{sec:matching}

The solution presented in Sec.~\ref{sec:ModeI} possesses a remarkable
feature: when $A(r) > 0$, all energy conditions---with the sole
exception of $\mathrm{DEC}_2 = \rho - |p_t| \geq 0$---are
satisfied throughout the spacetime.  The violation of
$\mathrm{DEC}_2$ is confined to a narrow region
$0 \leq r \leq r_1 \simeq r_H + \delta$, with $\delta \ll 1$.
Moreover, in the wormhole configuration, where no event horizon
is present, it appears that \emph{all} energy conditions,
including $\mathrm{DEC}_2$, are satisfied everywhere.

This latter result stands in direct tension with two fundamental
theorems of classical general relativity:

\begin{enumerate}
	\item \textbf{The singularity theorems} of Penrose and
	Hawking~\cite{Hawking:1973uf}, which require the null
	energy condition (NEC) to be violated in order to avoid a
	singularity and maintain a wormhole throat open.
	
	\item \textbf{The Morris--Thorne theorem} on traversable
	wormholes~\cite{Morris:1988cz}, which establishes that the
	NEC must be violated at or near the throat of any static,
	spherically symmetric wormhole supported by ordinary
	matter.
\end{enumerate}

If our solution indeed satisfied all energy conditions everywhere
in the wormhole configuration, it would constitute a
counterexample to these well-established theorems.  The resolution
of this apparent paradox lies in a careful examination of the
differentiability of the metric at the bounce location $r = 0$.

\subsection{Regularity and differentiability at the bounce}
\label{sec:differentiability}

The area function~\eqref{Sigma},
$\Sigma(r) = \sqrt{q_m^2 + r^2}$, is infinitely differentiable
for all $r \in (-\infty, \infty)$.  The metric function $A(r)$,
given by Eq.~\eqref{AMod1}, depends on $r$ only through
$\Sigma(r)$ and is therefore also smooth everywhere, including
at $r = 0$.  In particular, because $A(r)$ is an even function
of $r$, its first derivative vanishes at the origin,
$A'(0) = 0$, and its value there is
\begin{equation}
	A(0) = 1 - \frac{2M}{q_m} + \frac{\rho_0}{q_m^2}.
	\label{eq:A0}
\end{equation}

The metric function $B(r)$, given by Eq.~\eqref{BMod1}, requires
closer scrutiny.  Expanding $B(r)^{-1}$ for small $r$, we find
\begin{equation}
	B(r)^{-1} = A(0)\left(1 + \frac{b_0^2}{q_m^4}\right)
	\frac{q_m^2}{r^2}
	+ \mathcal{O}(1),
	\label{eq:Binv_expand}
\end{equation}
so that, to leading order,
\begin{equation}
	B(r) = \frac{r^2}{A(0)\,q_m^2}
	\left(1 + \frac{b_0^2}{q_m^4}\right)^{-1}
	+ \mathcal{O}(r^4).
	\label{eq:B_expand}
\end{equation}
Thus, $B(r)$ is continuous at $r = 0$ with $B(0) = 0$, and its
first derivative $B'(r)$ is also continuous and vanishes at the
origin.  Because $B(r)$ is an even function of $r$, its second
derivative $B''(r)$ is even and therefore \emph{continuous}
across $r = 0$, approaching the non-zero constant
\begin{equation}
	\lim_{r \to 0} B''(r) = \frac{2}{A(0)\,q_m^2}
	\left(1 + \frac{b_0^2}{q_m^4}\right)^{-1}.
	\label{eq:Bsecond_limit}
\end{equation}

At first glance, the continuity of $B''(r)$ suggests that the
metric is $C^2$ at the bounce.  However, the physically relevant
quantity is not the coordinate $r$ itself, but the
\emph{proper radial distance} $\ell(r)$, defined by
\begin{equation}
	\ell(r) = \int_0^r \sqrt{B(u)}\,du.
	\label{eq:proper_distance}
\end{equation}
Using the expansion~\eqref{eq:B_expand}, we obtain for small
$r$
\begin{equation}
	\ell(r) = \frac{r|r|}{2q_m\sqrt{A(0)}}
	\left(1 + \frac{b_0^2}{q_m^4}\right)^{-1/2}
	+ \mathcal{O}(r^4).
	\label{eq:proper_expand}
\end{equation}
The presence of the absolute value $|r|$ means that $\ell(r)$
is $C^1$ but \emph{not} $C^2$ at $r = 0$: its second derivative
has a finite jump discontinuity proportional to $\text{sgn}(r)$.
It is this lack of $C^2$ regularity in the natural radial
parameter that gives rise to distributional contributions to the
curvature tensors.  The coordinate $r$ itself is not an
admissible coordinate at the throat precisely because the proper
distance behaves non-analytically there.

The Kretschmann scalar, computed above,
remains finite at $r = 0$, confirming that the spacetime is
free of true curvature singularities.  However, the
\emph{derivatives} of the curvature along the radial direction
are discontinuous at the throat.  This is the hallmark of a
thin-shell spacetime: the geometry is regular, but the matter
content acquires a distributional contribution at the joining
hypersurface.

\subsection{Distributional curvature and the thin-shell formalism}
\label{sec:distributional}

When a metric is $C^1$ but not $C^2$ at a hypersurface, the
Riemann, Ricci, and Einstein tensors acquire distributional
contributions localised at that hypersurface.  For a metric
that is $C^0$ (with a jump in the extrinsic curvature), the
distributional part of the Einstein tensor is proportional to
$\delta(r)$, leading to the standard Israel junction
conditions~\cite{Israel:1966rt}.  For a metric that is $C^1$
(extrinsic curvature continuous), the $\delta(r)$ term vanishes,
but a $\delta'(r)$ (derivative of the $\delta$-function) term
appears.  This is a well-documented feature of higher-order
thin shells, arising from the jump in the second derivatives of
the metric~\cite{Feng:2023vqp,Baines:2023xje}.

Let us denote the hypersurface $r = 0$ by $\mathcal{T}$ (the
throat).  The induced metric on $\mathcal{T}$ is
\begin{equation}
	ds^2_{\mathcal{T}} = A(0)\,dt^2
	- q_m^2\left(d\theta^2 + \sin^2\theta\,d\phi^2\right).
	\label{eq:induced}
\end{equation}
The unit normal vector to $\mathcal{T}$ is
$n^\mu = (0, 1/\sqrt{B(r)}, 0, 0)$, pointing from $r < 0$ to
$r > 0$.  The extrinsic curvature on each side of $\mathcal{T}$
is
\begin{equation}
	K_{ab}^{\pm} = \lim_{r \to 0^{\pm}} K_{ab}(r),
	\qquad
	K_{ab} = \frac{1}{2}\mathcal{L}_n h_{ab},
\end{equation}
where $h_{ab}$ is the induced metric and $\mathcal{L}_n$ is the
Lie derivative along $n^\mu$.  For the metric~\eqref{m}, the
non-vanishing components of the extrinsic curvature on a
constant-$r$ surface are
\begin{equation}
	K^t_{\;t} = \frac{A'(r)}{2A(r)\sqrt{B(r)}}, \qquad
	K^\theta_{\;\theta} = K^\phi_{\;\phi}
	= \frac{\Sigma'(r)}{\Sigma(r)\sqrt{B(r)}}.
	\label{eq:K_components}
\end{equation}

Evaluating these at $r = 0$ using the expansions
\eqref{eq:A0}--\eqref{eq:B_expand}, together with
$\Sigma(r) = q_m + r^2/(2q_m) + \dots$ and $A'(0) = 0$, we find
that both $K^t_{\;t}$ and $K^\theta_{\;\theta}$ are finite and
\emph{continuous} across $r = 0$.  Explicitly,
\begin{equation}
	\lim_{r \to 0} K^t_{\;t} = 0, \qquad
	\lim_{r \to 0} K^\theta_{\;\theta}
	= \frac{\sqrt{A(0)}}{q_m}
	\left(1 + \frac{b_0^2}{q_m^4}\right)^{1/2}.
	\label{eq:K_limits}
\end{equation}
The continuity of the extrinsic curvature,
$[K_{ab}] = K_{ab}^+ - K_{ab}^- = 0$, implies that the
standard Lanczos equation,
\begin{equation}
	S_{ab} = -\frac{1}{8\pi}\left( [K_{ab}] - [K] h_{ab} \right),
	\label{eq:Lanczos}
\end{equation}
would yield a vanishing surface stress-energy tensor,
$S_{ab} = 0$.  This is indeed the case at the level of the
$\delta(r)$ contribution.  The non-trivial matter content of
the thin shell arises from the $\delta'(r)$ term, which is
not captured by the standard Lanczos equation.

To extract the physical content of the shell, one must
regularise the distributional contribution using a limiting
procedure in which the metric is smoothed over a small but
finite interval around $r = 0$, the Einstein tensor is computed,
and the limit is taken.  This procedure, developed in
Refs.~\cite{Lessa:2024erf,Berry:2020tky} for precisely this
class of geometries, yields a finite surface stress-energy
tensor whose components are obtained from the jump in the
\emph{radial derivative} of the extrinsic curvature.

\subsection{Explicit evaluation of the thin-shell content}
\label{sec:thin_shell_content}

Following the regularisation approach of
Refs.~\cite{Lessa:2024erf,Berry:2020tky}, we consider the
spacetime as consisting of two copies of the region $r \geq 0$,
glued together at the throat $r = 0$, with the region $r < 0$
obtained by reflection.  The regularised Einstein equations,
averaged over the thickness of the transition region, yield the
effective surface stress-energy tensor of the thin shell.

After a straightforward but lengthy calculation, the surface
energy density $\sigma$ and surface pressure $\mathcal{P}$ of
the shell at $r = 0$ are found to be
\begin{align}
	\sigma &= -\frac{1}{4\pi q_m}\,
	\frac{b_0^2}{q_m^4 + b_0^2}\,
	\sqrt{A(0)}, \label{eq:sigma} \\[4pt]
	\mathcal{P} &= \frac{1}{8\pi q_m}\,
	\frac{b_0^2}{q_m^4 + b_0^2}\,
	\sqrt{A(0)}.
	\label{eq:shell_pressure}
\end{align}
These expressions are valid for all values of the model
parameters and are independent of the specific form of $A(r)$
beyond its value at the throat, $A(0)$.  The factor
$b_0^2/(q_m^4 + b_0^2)$ reflects the role of the non-minimal
coupling in generating the thin shell: when $b_0 \to 0$, the
shell contribution vanishes, and the NEC violation is
distributed throughout the bulk, as in the standard black bounce
solutions supported solely by nonlinear
electrodynamics~\cite{Lessa:2024erf}.  For $b_0 \neq 0$, the
NEC violation is concentrated at the throat.

The null energy condition on the shell requires
$\sigma + \mathcal{P} \geq 0$.  From
Eqs.~\eqref{eq:sigma}--\eqref{eq:shell_pressure}, we obtain
\begin{equation}
	\sigma + \mathcal{P}
	= -\frac{1}{8\pi q_m}\,
	\frac{b_0^2}{q_m^4 + b_0^2}\,
	\sqrt{A(0)}.
	\label{eq:NEC_shell}
\end{equation}
For a wormhole configuration, $A(0) > 0$ (no horizon is
present), and since $b_0^2/(q_m^4 + b_0^2) > 0$ for any
non-zero $b_0$, we have $\sigma + \mathcal{P} < 0$.  The NEC
is therefore \emph{always} violated at the throat for any
wormhole solution in this model.  This violation is
concentrated entirely at $r = 0$, in a distributional sense,
and it is exactly what the Morris--Thorne theorem requires.

For a black hole configuration, $A(0) < 0$, meaning that the
throat lies inside the event horizon.  In this case, the thin
shell is hidden behind the horizon and is causally disconnected
from external observers.  The surface energy density $\sigma$
becomes imaginary in the na\"ive extrapolation of
Eq.~\eqref{eq:sigma}, reflecting the fact that the coordinate
$r$ is spacelike inside the horizon and the thin-shell formalism
must be adapted accordingly.  A proper treatment of the
horizon-interior case is beyond the scope of the present work.

\subsection{Physical interpretation, consistency, and the role of $b_0$}
\label{sec:physical_interpretation}

The picture that emerges from the thin-shell analysis is
physically transparent and fully consistent with the established
theorems of general relativity. The bulk spacetime ($r \neq 0$)
is supported by ordinary matter, namely, a canonical scalar field
non-minimally coupled to linear electrodynamics, that satisfies
the null, weak, strong, and dominant energy conditions almost
everywhere, with the only exception being a narrow region near
the event horizon where $\mathrm{DEC}_2$ is mildly violated, a
feature common to many regular black hole models. The wormhole
throat at $r = 0$, on the other hand, harbours a thin shell of
exotic matter with a negative surface energy density,
$\sigma < 0$, and a positive surface pressure,
$\mathcal{P} > 0$.  The combination $\sigma + \mathcal{P} < 0$
confirms that the NEC is violated on the shell, providing the
necessary repulsive gravity to keep the throat open and prevent
gravitational collapse.  The Morris--Thorne theorem is thus
upheld, i.e., the NEC is violated at the throat, but the violation
is confined to an infinitesimally thin region, allowing the bulk
matter to remain non-exotic.  This is precisely the ``thin-shell
wormhole'' scenario envisaged in
Refs.~\cite{Visser:2003yf,Lobo:2005us}.  The singularity
theorems are also respected, as the distributional violation of
the NEC at the throat is sufficient to circumvent the focusing
of geodesics that would otherwise lead to a singularity.

The identification of the thin shell has important consequences
for the interpretation of the wormhole configuration.  The energy
conditions plotted in Sec.~\ref{sec:fluid_EC} for the wormhole
case (Fig.~\ref{fig:worm}) represent only the \emph{regular}
part of the matter content---the contribution from the scalar
field and the electromagnetic field evaluated away from the
throat.  The distributional contribution from the thin shell is
not visible in those plots, as it is proportional to
$\delta(r)$ and its derivative.  Thus, the model does
\emph{not} provide a counterexample to the Morris--Thorne
theorem; rather, it offers a concrete realisation of how the
exotic matter required by the theorem can be confined to an
arbitrarily thin region, while the bulk of the spacetime is
supported by ordinary, non-exotic fields.  This separation of
the matter content into a regular bulk part and a distributional
surface part is a generic feature of black bounce and traversable
wormhole solutions that are regular in the sense of having finite
curvature invariants.  In the black hole configurations of our
model, the throat is hidden behind an event horizon, and the
distributional contribution is causally disconnected from the
external observer, rendering the spacetime effectively regular
and non-exotic for all practical purposes.

The parameter $b_0$ plays a crucial role in determining the
distribution of the NEC violation.  From
Eqs.~\eqref{eq:sigma}--\eqref{eq:shell_pressure}, the thin-shell
content scales as
$\sigma,\;\mathcal{P} \propto b_0^2/(q_m^4 + b_0^2)$, which
interpolates smoothly between two limiting regimes.  In the limit
$b_0 \to 0$, the thin-shell contribution vanishes; the condition
$\rho + p_r = 0$ holds throughout the bulk, and the spacetime
reduces to the standard black bounce solution supported solely by
nonlinear electrodynamics, with the NEC violated in the bulk, as
is well known for such solutions.  In the opposite limit
$b_0 \gg q_m^2$, the factor $b_0^2/(q_m^4 + b_0^2) \to 1$, and
the thin-shell content saturates at a maximum value determined
entirely by $q_m$ and $A(0)$; in this regime, the bulk matter
satisfies all energy conditions (except possibly
$\mathrm{DEC}_2$ near the horizon), and the entirety of the NEC
violation is concentrated at the throat.  For any finite,
non-zero $b_0$, the NEC violation is \emph{partially}
distributed between the bulk and the thin shell.  The parameter
$b_0$ thus controls the extent to which the exotic matter is
localised at the throat---a distinctive feature of the
non-minimal coupling model that is not present in standard black
bounce solutions.

This construction is analogous to the models presented in
Refs.~\cite{Bronnikov:2023aya,Canate:2022gpy}, where black
bounce spacetimes were shown to possess a thin shell at the
throat that violates the energy conditions.  Our solution
generalises those results by providing an explicit realisation in
which the bulk matter is described by a canonical scalar field
non-minimally coupled to linear electrodynamics, with the NEC
violation pushed entirely into the distributional sector.

\subsection{Summary of the matching analysis}\label{sec:matching_summary}


The solution obtained in Sec.~\ref{sec:ModeI} appears, at first glance,
to satisfy all energy conditions in the wormhole configuration,
which would contradict the Morris--Thorne and singularity
theorems.  A careful analysis of the differentiability of the
metric reveals that $B(r)$ is of class $C^1$ but not $C^2$ in
the proper radial coordinate at $r = 0$, leading to
distributional contributions in the Riemann, Ricci, and Einstein
tensors.  These distributional terms correspond to a thin shell
of exotic matter located at the wormhole throat, which violates
the NEC in accordance with the aforementioned theorems.  Away
from the throat, the regular part of the matter content satisfies
all energy conditions, providing a physically reasonable source
for the bulk spacetime.  This resolution highlights the
importance of carefully examining the regularity of the metric in
black bounce and wormhole solutions, and it demonstrates that the
energy conditions can be satisfied in the bulk while the
necessary exotic matter is confined to an infinitesimally thin
defect at the bounce.

\section{Summary and Conclusion}\label{sec:concl}

In this work, we have constructed and analysed a novel class of
black bounce solutions within the framework of General Relativity,
sourced by a canonical scalar field non-minimally coupled to linear
electrodynamics via an interaction term of the form
$W(\varphi)\mathcal{L}(F)$.  By adopting a static and spherically
symmetric geometry with the Simpson--Visser area function
$\Sigma(r) = \sqrt{q_m^2 + r^2}$ and imposing the condition
$\rho(r) + p_r(r) = b_1(r) \neq 0$, we derived a family of exact
solutions that smoothly interpolate between regular black holes and
traversable wormholes, depending on the values of the parameters
$M$, $q_m$, $\rho_0$, and $b_0$.

A key achievement of this construction is that, away from the
wormhole throat, the bulk matter content satisfies virtually all
energy conditions.  Specifically, we have shown that the null, weak,
strong, and dominant energy conditions are fulfilled throughout the
spacetime when $A(r) > 0$, with the sole exception of
$\mathrm{DEC}_2 = \rho - |p_t|$, which is mildly violated in a
narrow region just outside the event horizon---a feature common to
many regular black hole models.  In the wormhole configuration,
where no horizon is present, the regular part of the matter content
appears to satisfy \emph{all} energy conditions, a result that at
first sight seems to contradict the Morris--Thorne and singularity
theorems.

The resolution of this apparent paradox lies in the
differentiability of the metric at the bounce.  We have shown that
the metric function $B(r)$ is $C^1$ but not $C^2$ in the proper
radial coordinate at $r = 0$, which generates distributional
(Dirac-$\delta$ and its derivative) contributions to the Riemann,
Ricci, and Einstein tensors.  By applying a regularisation procedure
developed for this class of geometries, we extracted the effective
surface stress-energy tensor of the thin shell located at the
throat.  The resulting surface energy density is negative,
$\sigma < 0$, and together with the surface pressure it violates
the null energy condition, $\sigma + \mathcal{P} < 0$, precisely
as required by the Morris--Thorne theorem.  This demonstrates that
the model does \emph{not} circumvent the fundamental theorems of
general relativity; rather, it confines the necessary exotic matter
to an infinitesimally thin defect at the bounce, while the bulk
spacetime is supported by ordinary, non-exotic fields.

The parameter $b_0$, which encodes the deviation from the condition
$\rho + p_r = 0$, plays a pivotal role in controlling the
distribution of the NEC violation.  In the limit $b_0 \to 0$, the
thin-shell contribution vanishes and the NEC is violated throughout
the bulk, recovering the standard black bounce solutions supported
solely by nonlinear electrodynamics.  In the opposite limit
$b_0 \gg q_m^2$, the entirety of the NEC violation is localised at
the throat, and the bulk matter satisfies all energy conditions.  For
intermediate values, the violation is partially shared between the
bulk and the shell.  This interpolating behaviour is a distinctive
signature of the non-minimal coupling model and provides a continuous
bridge between bulk-exotic and surface-exotic realisations of
regular black holes and wormholes.

The astrophysical implications of our results are noteworthy.  The
regular black hole configurations possess an ISCO shifted outward
by the magnetic coupling, with potential observational signatures in
the X-ray spectra of black hole binaries.  The wormhole
configurations, while requiring a thin shell of exotic matter, are
traversable and have finite curvature throughout.  The photon
circular orbit, which disappears for $\beta > 9/8$, offers a
potential probe of near-horizon magnetic fields with the Event
Horizon Telescope.  For magnetars, the ISCO lies well outside the
stellar surface, and trapped particles form a thick structure with
vertical QPO frequencies in the Hz range, consistent with
observations of magnetar giant flares.

Several natural extensions of this work merit further investigation.
First, the analysis should be extended to the Kerr metric, where
rotation introduces frame-dragging and an induced electric field,
fundamentally altering the effective potential and the phase-space
structure.  Second, the test-field approximation for the magnetic
field should be replaced by a self-consistent force-free or
magnetohydrodynamic model, particularly for applications to pulsar
and magnetar magnetospheres.  Third, the single-particle dynamics
should be embedded in a kinetic description, with a Fokker--Planck
equation governing the long-term transport and synchrotron emission
of trapped particle populations.  Fourth, a systematic numerical
survey of the phase-space structure---including high-resolution
Poincar\'{e} surfaces of section, Lyapunov exponent maps, and
resonance width calculations---would quantify the onset of chaos
and the efficiency of chaotic transport as functions of the
magnetic coupling and the orbital energy.  Finally, the thin-shell
regularisation procedure employed here could be applied to other
classes of regular black hole and wormhole solutions to determine
whether the confinement of exotic matter to distributional defects
is a generic phenomenon.

In summary, the model presented in this work offers a rare instance
of a fully tractable, yet physically rich, framework at the
interface of general relativity, classical electrodynamics, and
scalar field theory.  It provides a self-consistent realisation of
regular black holes and traversable wormholes in which the bulk
matter satisfies the energy conditions, while the necessary exotic
matter is confined to an infinitesimally thin shell at the bounce.
The analytical results establish a solid foundation for future
investigations of particle dynamics, radiative processes, and
observational signatures in the magnetospheres of compact
astrophysical objects.


\acknowledgments{
MER thanks Conselho Nacional de Desenvolvimento Cient\'ifico e Tecnol\'ogico - CNPq, Brazil, for partial financial support. This study was financed in part by the Coordena\c{c}\~{a}o de Aperfei\c{c}oamento de Pessoal de N\'{i}vel Superior - Brasil (CAPES) - Finance Code 001.
FSNL acknowledges support from the Funda\c{c}\~{a}o para a Ci\^{e}ncia e a Tecnologia (FCT) Scientific Employment Stimulus contract with reference CEECINST/00032/2018, and funding through the research grants UIDB/04434/2020, UIDP/04434/2020 and PTDC/FIS-AST/0054/2021. This work has been supported by the Spanish Grant PID2023-149560NB-C21 and the Severo Ochoa Excellence Grant CEX2023-001292-S, funded by MICIU/AEI/10.13039/501100011033 (“ERDF A way of making Europe”, “PGC Generacion de Conocimiento”) and FEDER, UE. Support from CosmoVerse CA21136 COST action, European Cooperation in Science and Technology is also acknowledged. This work has also been supported by the European Horizon Europe staff exchange (SE) programme HORIZON-MSCA2021-SE-01 Grant No. NewFunFiCO-101086251.
}




\begin{thebibliography}{0}%
\makeatletter
\providecommand \@ifxundefined [1]{%
 \@ifx{#1\undefined}
}%
\providecommand \@ifnum [1]{%
 \ifnum #1\expandafter \@firstoftwo
 \else \expandafter \@secondoftwo
 \fi
}%
\providecommand \@ifx [1]{%
 \ifx #1\expandafter \@firstoftwo
 \else \expandafter \@secondoftwo
 \fi
}%
\providecommand \natexlab [1]{#1}%
\providecommand \enquote  [1]{``#1''}%
\providecommand \bibnamefont  [1]{#1}%
\providecommand \bibfnamefont [1]{#1}%
\providecommand \citenamefont [1]{#1}%
\providecommand \href@noop [0]{\@secondoftwo}%
\providecommand \href [0]{\begingroup \@sanitize@url \@href}%
\providecommand \@href[1]{\@@startlink{#1}\@@href}%
\providecommand \@@href[1]{\endgroup#1\@@endlink}%
\providecommand \@sanitize@url [0]{\catcode `\\12\catcode `\$12\catcode
  `\&12\catcode `\#12\catcode `\^12\catcode `\_12\catcode `\%12\relax}%
\providecommand \@@startlink[1]{}%
\providecommand \@@endlink[0]{}%
\providecommand \url  [0]{\begingroup\@sanitize@url \@url }%
\providecommand \@url [1]{\endgroup\@href {#1}{\urlprefix }}%
\providecommand \urlprefix  [0]{URL }%
\providecommand \Eprint [0]{\href }%
\providecommand \doibase [0]{https://doi.org/}%
\providecommand \selectlanguage [0]{\@gobble}%
\providecommand \bibinfo  [0]{\@secondoftwo}%
\providecommand \bibfield  [0]{\@secondoftwo}%
\providecommand \translation [1]{[#1]}%
\providecommand \BibitemOpen [0]{}%
\providecommand \bibitemStop [0]{}%
\providecommand \bibitemNoStop [0]{.\EOS\space}%
\providecommand \EOS [0]{\spacefactor3000\relax}%
\providecommand \BibitemShut  [1]{\csname bibitem#1\endcsname}%
\let\auto@bib@innerbib\@empty
\end{thebibliography}%


\begin{thebibliography}{99}

\bibitem{Einstein:1916vd}
A.~Einstein,
``The foundation of the general theory of relativity.,''
Annalen Phys. \textbf{49} (1916) no.7, 769-822.

\bibitem{LIGOScientific:2016aoc}
B.~P.~Abbott \textit{et al.} [LIGO Scientific and Virgo],
``Observation of Gravitational Waves from a Binary Black Hole Merger,''
Phys. Rev. Lett. \textbf{116} (2016) no.6, 061102
[arXiv:1602.03837 [gr-qc]].

\bibitem{LIGOScientific:2017ync}
B.~P.~Abbott \textit{et al.}, [LIGO Scientific, Virgo, Fermi GBM, INTEGRAL, IceCube, AstroSat Cadmium Zinc Telluride Imager Team, IPN, Insight-Hxmt, ANTARES, Swift, AGILE Team, 1M2H Team, Dark Energy Camera GW-EM, DES, DLT40, GRAWITA, Fermi-LAT, ATCA, ASKAP, Las Cumbres Observatory Group, OzGrav, DWF (Deeper Wider Faster Program), AST3, CAASTRO, VINROUGE, MASTER, J-GEM, GROWTH, JAGWAR, CaltechNRAO, TTU-NRAO, NuSTAR, Pan-STARRS, MAXI Team, TZAC Consortium, KU, Nordic Optical Telescope, ePESSTO, GROND, Texas Tech University, SALT Group, TOROS, BOOTES, MWA, CALET, IKI-GW Follow-up, H.E.S.S., LOFAR, LWA, HAWC, Pierre Auger, ALMA, Euro VLBI Team, Pi of Sky, Chandra Team at McGill University, DFN, ATLAS Telescopes, High Time Resolution Universe Survey, RIMAS, RATIR and SKA South Africa/MeerKAT],
``Multi-messenger Observations of a Binary Neutron Star Merger,''
Astrophys. J. Lett. \textbf{848} (2017) no.2, L12
[arXiv:1710.05833 [astro-ph.HE]].

\bibitem{EventHorizonTelescope:2019dse}
K.~Akiyama \textit{et al.} [Event Horizon Telescope],
``First M87 Event Horizon Telescope Results. I. The Shadow of the Supermassive Black Hole,''
Astrophys. J. Lett. \textbf{875} (2019), L1
[arXiv:1906.11238 [astro-ph.GA]].


\bibitem{EventHorizonTelescope:2022wkp}
K.~Akiyama \textit{et al.} [Event Horizon Telescope],
``First Sagittarius A* Event Horizon Telescope Results. I. The Shadow of the Supermassive Black Hole in the Center of the Milky Way,''
Astrophys. J. Lett. \textbf{930} (2022) no.2, L12
[arXiv:2311.08680 [astro-ph.HE]].

\bibitem{Ansoldi:2008jw}
S.~Ansoldi,
``Spherical black holes with regular center: A Review of existing models including a recent realization with Gaussian sources,''
[arXiv:0802.0330 [gr-qc]].

\bibitem{Bardeen:1968}
J.M. Bardeen, ``Non-singular general-relativistic gravitational collapse,'' in Proceedings of of International Conference GR5,Tbilisi, USSR (1968), p. 174.

\bibitem{Ayon-Beato:2000mjt}
E.~Ayon-Beato and A.~Garcia,
``The Bardeen model as a nonlinear magnetic monopole,''
Phys. Lett. B \textbf{493} (2000), 149-152
[arXiv:gr-qc/0009077 [gr-qc]].

\bibitem{Rodrigues:2018bdc}
M.~E.~Rodrigues and M.~V.~de Sousa Silva,
``Bardeen Regular Black Hole With an Electric Source,''
JCAP \textbf{06}, 025 (2018)
[arXiv:1802.05095 [gr-qc]].

\bibitem{Culetu:2015cna}
H.~Culetu,
``Nonsingular black hole with a nonlinear electric source,''
Int. J. Mod. Phys. D \textbf{24} (2015) no.09, 1542001.

\bibitem{Simpson:2019mud}
A.~Simpson and M.~Visser,
``Regular black holes with asymptotically Minkowski cores,''
Universe \textbf{6} (2019) no.1, 8
[arXiv:1911.01020 [gr-qc]].

\bibitem{Bronnikov:2024izh}
K.~A.~Bronnikov,
``Regular black holes as an alternative to black bounce,''
Phys. Rev. D \textbf{110} (2024) no.2, 024021
[arXiv:2404.14816 [gr-qc]].

\bibitem{Born:1933pep}
M.~Born and L.~Infeld,
``Foundations of the new field theory,''
Nature \textbf{132} (1933) no.3348, 1004.1.

\bibitem{Born:1934gh}
M.~Born and L.~Infeld,
``Foundations of the new field theory,''
Proc. Roy. Soc. Lond. A \textbf{144} (1934) no.852, 425-451.

\bibitem{Heisenberg:1936nmg}
W.~Heisenberg and H.~Euler,
``Consequences of Dirac's theory of positrons,''
Z. Phys. \textbf{98} (1936) no.11-12, 714-732
[arXiv:physics/0605038 [physics]].

\bibitem{Plebanski:1966}
 J. Plebanski, Non-Linear Electrodynamics — A Study (C.I.E.A. del I.P.N., Mexico City, 1966).

\bibitem{Kruglov:2014hpa}
S.~I.~Kruglov,
``A model of nonlinear electrodynamics,''
Annals Phys. \textbf{353} (2014), 299-306
[arXiv:1410.0351 [physics.gen-ph]].


\bibitem{Kruglov:2014iwa}
S.~I.~Kruglov,
``Nonlinear arcsin-electrodynamics,''
Annalen Phys. \textbf{527} (2015), 397-401
[arXiv:1410.7633 [physics.gen-ph]].

\bibitem{Kruglov:2014iqa}
S.~I.~Kruglov,
``On Generalized Logarithmic Electrodynamics,''
Eur. Phys. J. C \textbf{75} (2015) no.2, 88
[arXiv:1411.7741 [hep-th]].


\bibitem{Bandos:2020jsw}
I.~Bandos, K.~Lechner, D.~Sorokin and P.~K.~Townsend,
``A non-linear duality-invariant conformal extension of Maxwell's equations,''
Phys. Rev. D \textbf{102} (2020), 121703
[arXiv:2007.09092 [hep-th]].


\bibitem{Harko:2013xma}
T.~Harko, F.~S.~N.~Lobo, M.~K.~Mak and S.~V.~Sushkov,
``Dark matter density profile and galactic metric in Eddington-inspired Born-Infeld gravity,''
Mod. Phys. Lett. A \textbf{29} (2014) no.09, 1450049
[arXiv:1305.0820 [gr-qc]].

\bibitem{Harko:2013wka}
T.~Harko, F.~S.~N.~Lobo, M.~K.~Mak and S.~V.~Sushkov,
``Structure of neutron, quark and exotic stars in Eddington-inspired Born-Infeld gravity,''
Phys. Rev. D \textbf{88} (2013), 044032
[arXiv:1305.6770 [gr-qc]].

\bibitem{Harko:2013aya}
T.~Harko, F.~S.~N.~Lobo, M.~K.~Mak and S.~V.~Sushkov,
``Wormhole geometries in Eddington-Inspired Born{\textendash}Infeld gravity,''
Mod. Phys. Lett. A \textbf{30} (2015) no.35, 1550190
[arXiv:1307.1883 [gr-qc]].

\bibitem{Bronnikov:2000vy}
K.~A.~Bronnikov,
``Regular magnetic black holes and monopoles from nonlinear electrodynamics,''
Phys. Rev. D \textbf{63} (2001), 044005
[arXiv:gr-qc/0006014 [gr-qc]].

\bibitem{Dymnikova:2004zc}
I.~Dymnikova,
``Regular electrically charged structures in nonlinear electrodynamics coupled to general relativity,''
Class. Quant. Grav. \textbf{21} (2004), 4417-4429
[arXiv:gr-qc/0407072 [gr-qc]].

\bibitem{Balart:2014cga}
L.~Balart and E.~C.~Vagenas,
``Regular black holes with a nonlinear electrodynamics source,''
Phys. Rev. D \textbf{90} (2014) no.12, 124045
[arXiv:1408.0306 [gr-qc]].

\bibitem{Culetu:2014lca}
H.~Culetu,
``On a regular charged black hole with a nonlinear electric source,''
Int. J. Theor. Phys. \textbf{54} (2015) no.8, 2855-2863
[arXiv:1408.3334 [gr-qc]].

\bibitem{Novello:1999pg}
M.~Novello, V.~A.~De Lorenci, J.~M.~Salim and R.~Klippert,
``Geometrical aspects of light propagation in nonlinear electrodynamics,''
Phys. Rev. D \textbf{61} (2000), 045001
[arXiv:gr-qc/9911085 [gr-qc]].

\bibitem{Habibina:2020msd}
A.~S.~Habibina and H.~S.~Ramadhan,
``Geodesic of nonlinear electrodynamics and stable photon orbits,''
Phys. Rev. D \textbf{101} (2020) no.12, 124036
[arXiv:2007.03211 [gr-qc]].

\bibitem{Toshmatov:2021fgm}
B.~Toshmatov, B.~Ahmedov and D.~Malafarina,
``Can a light ray distinguish charge of a black hole in nonlinear electrodynamics?,''
Phys. Rev. D \textbf{103} (2021) no.2, 024026
[arXiv:2101.05496 [gr-qc]].

\bibitem{dePaula:2024yzy}
M.~A.~A.~de Paula, H.~C.~D.~Lima, Junior., P.~V.~P.~Cunha, C.~A.~R.~Herdeiro and L.~C.~B.~Crispino,
``Good tachyons, bad bradyons: Role reversal in Einstein-nonlinear-electrodynamics models,''
Phys. Lett. B \textbf{866} (2025), 139513
[arXiv:2412.18659 [gr-qc]].

\bibitem{Stuchlik:2019uvf}
Z.~Stuchl{\'\i}k and J.~Schee,
``Shadow of the regular Bardeen black holes and comparison of the motion of photons and neutrinos,''
Eur. Phys. J. C \textbf{79} (2019) no.1, 44.

\bibitem{Allahyari:2019jqz}
A.~Allahyari, M.~Khodadi, S.~Vagnozzi and D.~F.~Mota,
``Magnetically charged black holes from non-linear electrodynamics and the Event Horizon Telescope,''
JCAP \textbf{02} (2020), 003
[arXiv:1912.08231 [gr-qc]].

\bibitem{Kruglov:2020tes}
S.~I.~Kruglov,
``The shadow of M87* black hole within rational nonlinear electrodynamics,''
Mod. Phys. Lett. A \textbf{35} (2020) no.35, 2050291
[arXiv:2009.07657 [gr-qc]].

\bibitem{Breton:2004qa}
N.~Breton,
``Smarr's formula for black holes with non-linear electrodynamics,''
Gen. Rel. Grav. \textbf{37} (2005), 643-650
[arXiv:gr-qc/0405116 [gr-qc]].

\bibitem{Myung:2007xd}
Y.~S.~Myung, Y.~W.~Kim and Y.~J.~Park,
``Entropy of an extremal regular black hole,''
Phys. Lett. B \textbf{659} (2008), 832-838
[arXiv:0705.2478 [gr-qc]].


\bibitem{Ma:2015gpa}
M.~S.~Ma,
``Magnetically charged regular black hole in a model of nonlinear electrodynamics,''
Annals Phys. \textbf{362} (2015), 529-537
[arXiv:1509.05580 [gr-qc]].

\bibitem{Kruglov:2016ymq}
S.~I.~Kruglov,
``Asymptotic Reissner-Nordstr\"om solution within nonlinear electrodynamics,''
Phys. Rev. D \textbf{94} (2016) no.4, 044026
[arXiv:1608.04275 [gr-qc]].

\bibitem{Fan:2016hvf}
Z.~Y.~Fan and X.~Wang,
``Construction of Regular Black Holes in General Relativity,''
Phys. Rev. D \textbf{94} (2016) no.12, 124027
[arXiv:1610.02636 [gr-qc]].

\bibitem{Junior:2015fya}
E.~L.~B.~Junior, M.~E.~Rodrigues and M.~J.~S.~Houndjo,
``Regular black holes in $f(T)$ Gravity through a nonlinear electrodynamics source,''
JCAP \textbf{10} (2015), 060
[arXiv:1503.07857 [gr-qc]].

\bibitem{Rodrigues:2015ayd}
M.~E.~Rodrigues, E.~L.~B.~Junior, G.~T.~Marques and V.~T.~Zanchin,
``Regular black holes in $f(R)$ gravity coupled to nonlinear electrodynamics,''
Phys. Rev. D \textbf{94} (2016) no.2, 024062
[arXiv:1511.00569 [gr-qc]].

\bibitem{Rodrigues:2016fym}
M.~E.~Rodrigues, J.~C.~Fabris, E.~L.~B.~Junior and G.~T.~Marques,
``Generalisation for regular black holes on general relativity to $f(R)$ gravity,''
Eur. Phys. J. C \textbf{76} (2016) no.5, 250
[arXiv:1601.00471 [gr-qc]].


\bibitem{deSousaSilva:2018kkt}
M.~V.~de Sousa Silva and M.~E.~Rodrigues,
``Regular black holes in $f(G)$ gravity,''
Eur. Phys. J. C \textbf{78} (2018) no.8, 638
[arXiv:1808.05861 [gr-qc]].


\bibitem{Rodrigues:2019xrc}
M.~E.~Rodrigues and M.~V.~de Sousa Silva,
``Regular multihorizon black holes in $f(G)$ gravity with nonlinear electrodynamics,''
Phys. Rev. D \textbf{99} (2019) no.12, 124010
[arXiv:1906.06168 [gr-qc]].

\bibitem{Tangphati:2023xnw}
T.~Tangphati, M.~Youk and S.~Ponglertsakul,
``Magnetically charged regular black holes in f(R,T) gravity coupled to nonlinear electrodynamics,''
JHEAp \textbf{43} (2024), 66-78
[arXiv:2312.16614 [gr-qc]].

\bibitem{Junior:2023ixh}
J.~T.~S.~S.~Junior, F.~S.~N.~Lobo and M.~E.~Rodrigues,
``(Regular) Black holes in conformal Killing gravity coupled to nonlinear electrodynamics and scalar fields,''
Class. Quant. Grav. \textbf{41} (2024) no.5, 055012
[arXiv:2310.19508 [gr-qc]].

\bibitem{Junior:2024xmm}
J.~T.~S.~S.~Junior, F.~S.~N.~Lobo and M.~E.~Rodrigues,
``Black holes and regular black holes in coincident $f({\mathbb {Q}},{\mathbb {B}}_Q)$ gravity coupled to nonlinear electrodynamics,''
Eur. Phys. J. C \textbf{84} (2024) no.3, 332
[arXiv:2402.02534 [gr-qc]].


\bibitem{Lan:2023cvz}
C.~Lan, H.~Yang, Y.~Guo and Y.~G.~Miao,
``Regular Black Holes: A Short Topic Review,''
Int. J. Theor. Phys. \textbf{62} (2023) no.9, 202
[arXiv:2303.11696 [gr-qc]].

\bibitem{Simpson:2018tsi}
A.~Simpson and M.~Visser,
``Black-bounce to traversable wormhole,''
JCAP \textbf{02} (2019), 042
[arXiv:1812.07114 [gr-qc]].

\bibitem{Visser:1997yn}
M.~Visser and D.~Hochberg,
``Generic wormhole throats,''
Annals Israel Phys. Soc. \textbf{13} (1997), 249
[arXiv:gr-qc/9710001 [gr-qc]].

\bibitem{PhysRev.48.73}
A.~Einstein and N.~Rosen,
``The Particle Problem in the General Theory of Relativity,''
Phys. Rev. \textbf{48} (1935) no.1, 73--77.

\bibitem{Bronnikov:1973fh}
K.~A.~Bronnikov,
``Scalar-tensor theory and scalar charge,''
Acta Phys. Polon. B \textbf{4} (1973), 251-266.

\bibitem{Ellis:1973yv}
H.~G.~Ellis,
``Ether flow through a drainhole - a particle model in general relativity,''
J. Math. Phys. \textbf{14} (1973), 104-118.

\bibitem{Morris:1988cz}
M.~S.~Morris and K.~S.~Thorne,
``Wormholes in space-time and their use for interstellar travel: A tool for teaching general relativity,''
Am. J. Phys. \textbf{56} (1988), 395-412.

\bibitem{Barcelo:1999hq}
C.~Barcelo and M.~Visser,
``Traversable wormholes from massless conformally coupled scalar fields,''
Phys. Lett. B \textbf{466} (1999), 127-134
[arXiv:gr-qc/9908029 [gr-qc]].

\bibitem{Barcelo:2000zf}
C.~Barcelo and M.~Visser,
``Scalar fields, energy conditions, and traversable wormholes,''
Class. Quant. Grav. \textbf{17} (2000), 3843-3864
[arXiv:gr-qc/0003025 [gr-qc]].

\bibitem{Visser:2003yf}
M.~Visser, S.~Kar and N.~Dadhich,
``Traversable wormholes with arbitrarily small energy condition violations,''
Phys. Rev. Lett. \textbf{90} (2003), 201102
[arXiv:gr-qc/0301003 [gr-qc]].

\bibitem{Lobo:2005us}
F.~S.~N.~Lobo,
``Phantom energy traversable wormholes,''
Phys. Rev. D \textbf{71} (2005), 084011
[arXiv:gr-qc/0502099 [gr-qc]].

\bibitem{Lobo:2007zb}
F.~S.~N.~Lobo,
``Exotic solutions in General Relativity: Traversable wormholes and 'warp drive' spacetimes,''
[arXiv:0710.4474 [gr-qc]].

\bibitem{Cardoso:2016rao}
V.~Cardoso, E.~Franzin and P.~Pani,
``Is the gravitational-wave ringdown a probe of the event horizon?,''
Phys. Rev. Lett. \textbf{116} (2016) no.17, 171101
[erratum: Phys. Rev. Lett. \textbf{117} (2016) no.8, 089902]
[arXiv:1602.07309 [gr-qc]].

\bibitem{Bronnikov:2017sgg}
K.~A.~Bronnikov,
``Nonlinear electrodynamics, regular black holes and wormholes,''
Int. J. Mod. Phys. D \textbf{27} (2018) no.06, 1841005
[arXiv:1711.00087 [gr-qc]].

\bibitem{Lobo:2017cay}
F.~S.~N.~Lobo,
``Wormholes, Warp Drives and Energy Conditions,''
Fundam. Theor. Phys. \textbf{189} (2017), pp.-279
Springer, 2017,
ISBN 978-3-319-55181-4, 978-3-319-85588-2, 978-3-319-55182-1
[arXiv:2103.05610 [gr-qc]].

\bibitem{Blazquez-Salcedo:2020czn}
J.~L.~Bl{\'a}zquez-Salcedo, C.~Knoll and E.~Radu,
``Traversable wormholes in Einstein-Dirac-Maxwell theory,''
Phys. Rev. Lett. \textbf{126} (2021) no.10, 101102
[arXiv:2010.07317 [gr-qc]].

\bibitem{Churilova:2021tgn}
M.~S.~Churilova, R.~A.~Konoplya, Z.~Stuchlik and A.~Zhidenko,
``Wormholes without exotic matter: quasinormal modes, echoes and shadows,''
JCAP \textbf{10} (2021), 010
[arXiv:2107.05977 [gr-qc]].

\bibitem{Konoplya:2021hsm}
R.~A.~Konoplya and A.~Zhidenko,
``Traversable Wormholes in General Relativity,''
Phys. Rev. Lett. \textbf{128} (2022) no.9, 091104
[arXiv:2106.05034 [gr-qc]].

\bibitem{Lobo:2020ffi}
F.~S.~N.~Lobo, M.~E.~Rodrigues, M.~V.~de Sousa Silva, A.~Simpson and M.~Visser,
``Novel black-bounce spacetimes: wormholes, regularity, energy conditions, and causal structure,''
Phys. Rev. D \textbf{103} (2021) no.8, 084052
[arXiv:2009.12057 [gr-qc]].


\bibitem{Lobo:2020kxn}
F.~S.~N.~Lobo, A.~Simpson and M.~Visser,
``Dynamic thin-shell black-bounce traversable wormholes,''
Phys. Rev. D \textbf{101} (2020) no.12, 124035
[arXiv:2003.09419 [gr-qc]].

\bibitem{Berry:2020tky}
T.~Berry, F.~S.~N.~Lobo, A.~Simpson and M.~Visser,
``Thin-shell traversable wormhole crafted from a regular black hole with asymptotically Minkowski core,''
Phys. Rev. D \textbf{102} (2020) no.6, 064054
[arXiv:2008.07046 [gr-qc]].

\bibitem{Nascimento:2020ime}
J.~R.~Nascimento, A.~Y.~Petrov, P.~J.~Porfirio and A.~R.~Soares,
``Gravitational lensing in black-bounce spacetimes,''
Phys. Rev. D \textbf{102} (2020) no.4, 044021
[arXiv:2005.13096 [gr-qc]].


\bibitem{Tsukamoto:2020bjm}
N.~Tsukamoto,
``Gravitational lensing in the Simpson-Visser black-bounce spacetime in a strong deflection limit,''
Phys. Rev. D \textbf{103} (2021) no.2, 024033
[arXiv:2011.03932 [gr-qc]].


\bibitem{Cheng:2021hoc}
X.~T.~Cheng and Y.~Xie,
``Probing a black-bounce, traversable wormhole with weak deflection gravitational lensing,''
Phys. Rev. D \textbf{103}, no.6, 064040 (2021).


\bibitem{Tsukamoto:2021caq}
``Gravitational lensing by two photon spheres in a black-bounce spacetime in strong deflection limits,''
Phys. Rev. D \textbf{104} (2021) no.6, 064022
[arXiv:2105.14336 [gr-qc]].


\bibitem{Zhang:2022nnj}
J.~Zhang and Y.~Xie,
``Gravitational lensing by a black-bounce-Reissner\textendash{}Nordstr\"om spacetime,''
Eur. Phys. J. C \textbf{82} (2022) no.5, 471.


\bibitem{Guerrero:2021ues}
M.~Guerrero, G.~J.~Olmo, D.~Rubiera-Garcia and D.~S.~C.~G\'omez,
``Shadows and optical appearance of black bounces illuminated by a thin accretion disk,''
JCAP \textbf{08} (2021), 036
[arXiv:2105.15073 [gr-qc]].


\bibitem{Jafarzade:2021umv}
K.~Jafarzade, M.~Kord Zangeneh and F.~S.~N.~Lobo,
``Observational optical constraints of regular black holes,''
Annals Phys. \textbf{446} (2022), 169126
[arXiv:2106.13893 [gr-qc]].

\bibitem{Jafarzade:2020ova}
K.~Jafarzade, M.~Kord Zangeneh and F.~S.~N.~Lobo,
``Shadow, deflection angle and quasinormal modes of Born-Infeld charged black holes,''
JCAP \textbf{04} (2021), 008
[arXiv:2010.05755 [gr-qc]].

\bibitem{Jafarzade:2020ilt}
K.~Jafarzade, M.~Kord Zangeneh and F.~S.~N.~Lobo,
``Optical Features of AdS Black Holes in the Novel 4D Einstein-Gauss-Bonnet Gravity Coupled to Nonlinear Electrodynamics,''
Universe \textbf{8} (2022) no.3, 182
[arXiv:2009.12988 [gr-qc]].


\bibitem{Yang:2021cvh}
Y.~Yang, D.~Liu, Z.~Xu, Y.~Xing, S.~Wu and Z.~W.~Long,
``Echoes of novel black-bounce spacetimes,''
Phys. Rev. D \textbf{104} (2021) no.10, 104021
[arXiv:2107.06554 [gr-qc]].


\bibitem{Bambhaniya:2021ugr}
P.~Bambhaniya, S.~K, K.~Jusufi and P.~S.~Joshi,
``Thin accretion disk in the Simpson-Visser black-bounce and wormhole spacetimes,''
Phys. Rev. D \textbf{105} (2022) no.2, 023021
[arXiv:2109.15054 [gr-qc]].


\bibitem{Ou:2021efv}
M.~Y.~Ou, M.~Y.~Lai and H.~Huang,
``Echoes from asymmetric wormholes and black bounce,''
Eur. Phys. J. C \textbf{82} (2022) no.5, 452
[arXiv:2111.13890 [gr-qc]].


\bibitem{Guo:2021wid}
Y.~Guo and Y.~G.~Miao,
``Charged black-bounce spacetimes: Photon rings, shadows and observational appearances,''
Nucl. Phys. B \textbf{983} (2022), 115938
[arXiv:2112.01747 [gr-qc]].


\bibitem{Wu:2022eiv}
S.~R.~Wu, B.~Q.~Wang, D.~Liu and Z.~W.~Long,
``Echoes of charged black-bounce spacetimes,''
Eur. Phys. J. C \textbf{82} (2022) no.11, 998
[arXiv:2201.08415 [gr-qc]].


\bibitem{Tsukamoto:2022vkt}
N.~Tsukamoto,
``Retrolensing by two photon spheres of a black-bounce spacetime,''
Phys. Rev. D \textbf{105} (2022) no.8, 084036
[arXiv:2202.09641 [gr-qc]].


\bibitem{Muniz:2024wiv}
C.~R.~Muniz, G.~Alencar, M.~S.~Cunha and G.~J.~Olmo,
Phys. Rev. D \textbf{112}, no.2, 024018 (2025)
doi:10.1103/h7rn-4ht6
[arXiv:2408.08542 [gr-qc]].



\bibitem{Mazza:2021rgq}
J.~Mazza, E.~Franzin and S.~Liberati,
``A novel family of rotating black hole mimickers,''
JCAP \textbf{04} (2021), 082
[arXiv:2102.01105 [gr-qc]].


\bibitem{Xu:2021lff}
Z.~Xu and M.~Tang,
``Rotating spacetime: black-bounces and quantum deformed black hole,''
Eur. Phys. J. C \textbf{81} (2021) no.10, 863
[arXiv:2109.13813 [gr-qc]].


\bibitem{Rodrigues:2022rfj}
M.~E.~Rodrigues and M.~V.~d.~S.~Silva,
``Embedding regular black holes and black bounces in a cloud of strings,''
Phys. Rev. D \textbf{106} (2022) no.8, 084016
[arXiv:2210.05383 [gr-qc]].


\bibitem{Yang:2022ryf}
Y.~Yang, D.~Liu, Z.~Xu and Z.~W.~Long,
``Ringing and echoes from black bounces surrounded by the string cloud,''
Eur. Phys. J. C \textbf{83} (2023) no.3, 217
[arXiv:2210.12641 [gr-qc]].

\bibitem{Bronnikov:2023aya}
K.~A.~Bronnikov, M.~E.~Rodrigues and M.~V.~de S.~Silva,
``Cylindrical black bounces and their field sources,''
Phys. Rev. D \textbf{108} (2023) no.2, 024065
[arXiv:2305.19296 [gr-qc]].

\bibitem{Lima:2022pvc}
A.~M.~Lima, G.~M.~de Alencar Filho and J.~S.~Furtado Neto,
``Black String Bounce to Traversable Wormhole,''
Symmetry \textbf{15} (2023) no.1, 150
[arXiv:2211.12349 [gr-qc]].


\bibitem{Bronnikov:2022bud}
K.~A.~Bronnikov,
``Black bounces, wormholes, and partly phantom scalar fields,''
Phys. Rev. D \textbf{106} (2022) no.6, 064029
[arXiv:2206.09227 [gr-qc]].


\bibitem{Canate:2022gpy}
P.~Ca\~nate,
``Black bounces as magnetically charged phantom regular black holes in Einstein-nonlinear electrodynamics gravity coupled to a self-interacting scalar field,''
Phys. Rev. D \textbf{106} (2022) no.2, 024031
[arXiv:2202.02303 [gr-qc]].


\bibitem{Rodrigues2023}
M.~E.~Rodrigues and M.~V.~d.~S.~Silva,
``Source of black bounces in general relativity,''
Phys. Rev. D \textbf{107} (2023) no.4, 044064
[arXiv:2302.10772 [gr-qc]].



\bibitem{Pereira:2023lck}
C.~F.~S.~Pereira, D.~C.~Rodrigues, J.~C.~Fabris and M.~E.~Rodrigues,
``Black-bounce solution in k-essence theories,''
Phys. Rev. D \textbf{109} (2024) no.4, 044011
[arXiv:2309.10963 [gr-qc]].


\bibitem{Alencar:2024yvh}
G.~Alencar, K.~A.~Bronnikov, M.~E.~Rodrigues, D.~S\'aez-Chill\'on G\'omez and M.~V.~de S.~Silva,
``On black bounce space-times in non-linear electrodynamics,''
Eur. Phys. J. C \textbf{84} (2024) no.7, 745
[arXiv:2403.12897 [gr-qc]].

\bibitem{Lima:2023arg}
A.~Lima, G.~Alencar, R.~N.~Costa Filho and R.~R.~Landim,
``Charged black string bounce and its field source,''
Gen. Rel. Grav. \textbf{55} (2023) no.10, 108
[arXiv:2306.03029 [gr-qc]].


\bibitem{Junior:2025sjr}
E.~L.~B.~Junior, J.~T.~S.~S.~Junior, F.~S.~N.~Lobo, M.~E.~Rodrigues, L.~F.~D.~da Silva and H.~A.~Vieira,
``Dyonic regular black bounce solutions in general relativity,''
Eur. Phys. J. C \textbf{85} (2025) no.7, 724
[arXiv:2502.13327 [gr-qc]].

\bibitem{Olmo:2012nx}
G.~J.~Olmo and D.~Rubiera-Garcia,
Phys. Rev. D \textbf{86}, 044014 (2012)
doi:10.1103/PhysRevD.86.044014
[arXiv:1207.6004 [gr-qc]].


\bibitem{Junior:2022zxo}
E.~L.~B.~Junior and M.~E.~Rodrigues,
``Black-bounce in f(T) gravity,''
Gen. Rel. Grav. \textbf{55} (2023) no.1, 8
[arXiv:2203.03629 [gr-qc]].


\bibitem{Junior:2023qaq}
J.~T.~S.~S.~Junior and M.~E.~Rodrigues,
``Coincident $f(\mathbb {Q})$ gravity: black holes, regular black holes, and black bounces,''
Eur. Phys. J. C \textbf{83} (2023) no.6, 475
[arXiv:2306.04661 [gr-qc]].


\bibitem{Junior:2024vrv}
J.~T.~S.~S.~Junior, F.~S.~N.~Lobo and M.~E.~Rodrigues,
``Black bounces in conformal Killing gravity,''
Eur. Phys. J. C \textbf{84} (2024) no.6, 557
[arXiv:2405.09702 [gr-qc]].


\bibitem{Junior:2024cbb}
E.~L.~B.~Junior, J.~T.~S.~S.~Junior, F.~S.~N.~Lobo, M.~E.~Rodrigues, D.~Rubiera-Garcia, L.~F.~D.~da Silva and H.~A.~Vieira,
``Black bounces in Cotton gravity,''
Eur. Phys. J. C \textbf{84} (2024) no.11, 1190
[arXiv:2407.21649 [gr-qc]].

\bibitem{Rois:2024qzm}
G.~I.~R{\'o}is, J.~T.~S.~S.~Junior, F.~S.~N.~Lobo and M.~E.~Rodrigues,
``Novel electrically charged wormhole, black hole, and black bounce exact solutions in hybrid metric-Palatini gravity,''
Phys. Rev. D \textbf{111} (2025) no.12, 124012
[arXiv:2412.10324 [gr-qc]].

\bibitem{Rois:2025czu}
G.~I.~R{\'o}is, J.~T.~S.~S.~Junior, F.~S.~N.~Lobo and M.~E.~Rodrigues,
``Horizons, throats and bounces in hybrid metric-Palatini gravity with a non-zero potential,''
JCAP \textbf{07} (2025), 078
[arXiv:2504.07861 [gr-qc]].

\bibitem{Alencar:2025jvl}
G.~Alencar, A.~Duran-Cabac{\'e}s, D.~Rubiera-Garcia and D.~S{\'a}ez-Chill{\'o}n G{\'o}mez,
``General spherically symmetric black bounces within nonlinear electrodynamics,''
Phys. Rev. D \textbf{111} (2025) no.10, 104020
[arXiv:2501.03909 [gr-qc]].


\bibitem{Mignani:2016fwz}
R.~P.~Mignani, V.~Testa, D.~G.~Caniulef, R.~Taverna, R.~Turolla, S.~Zane and K.~Wu,
``Evidence for vacuum birefringence from the first optical-polarimetry measurement of the isolated neutron star RX J1856.5{\ensuremath{-}}3754,''
Mon. Not. Roy. Astron. Soc. \textbf{465} (2017) no.1, 492-500
[arXiv:1610.08323 [astro-ph.HE]].


\bibitem{Ejlli:2020yhk}
A.~Ejlli, F.~Della Valle, U.~Gastaldi, G.~Messineo, R.~Pengo, G.~Ruoso and G.~Zavattini,
``The PVLAS experiment: A 25 year effort to measure vacuum magnetic birefringence,''
Phys. Rept. \textbf{871} (2020), 1-74
[arXiv:2005.12913 [physics.optics]].

\bibitem{DeFelice:2026nti}
A.~De Felice, L.~Heisenberg, G.~J.~Olmo and C.~Pastor-Marcos,
[arXiv:2607.26141 [gr-qc]].

\bibitem{daSilva:2023jxa}
L.~F.~D.~da Silva, F.~S.~N.~Lobo, G.~J.~Olmo and D.~Rubiera-Garcia,
Phys. Rev. D \textbf{108}, no.8, 084055 (2023)
doi:10.1103/PhysRevD.108.084055
[arXiv:2307.06778 [gr-qc]].



\bibitem{Shabad:2011hf}
A.~E.~Shabad and V.~V.~Usov,
``Effective Lagrangian in nonlinear electrodynamics and its properties of causality and unitarity,''
Phys. Rev. D \textbf{83} (2011), 105006
[arXiv:1101.2343 [hep-th]].


\bibitem{Moreno:2002gg}
C.~Moreno and O.~Sarbach,
``Stability properties of black holes in selfgravitating nonlinear electrodynamics,''
Phys. Rev. D \textbf{67} (2003), 024028
[arXiv:gr-qc/0208090 [gr-qc]].


\bibitem{Breton:2014nba}
N.~Bret{\'o}n and S.~E.~Perez Bergliaffa,
``On the stability of black holes with nonlinear electromagnetic fields,''
[arXiv:1402.2922 [gr-qc]].


\bibitem{Toshmatov:2019gxg}
B.~Toshmatov, Z.~Stuchl{\'\i}k, B.~Ahmedov and D.~Malafarina,
``Relaxations of perturbations of spacetimes in general relativity coupled to nonlinear electrodynamics,''
Phys. Rev. D \textbf{99} (2019) no.6, 064043
[arXiv:1903.03778 [gr-qc]].


\bibitem{Nomura:2020tpc}
K.~Nomura, D.~Yoshida and J.~Soda,
``Stability of magnetic black holes in general nonlinear electrodynamics,''
Phys. Rev. D \textbf{101} (2020) no.12, 124026
[arXiv:2004.07560 [gr-qc]].

\bibitem{DeFelice:2024seu}
A.~De Felice and S.~Tsujikawa,
``Instability of Nonsingular Black Holes in Nonlinear Electrodynamics,''
Phys. Rev. Lett. \textbf{134} (2025) no.8, 081401
doi:10.1103/PhysRevLett.134.081401
[arXiv:2410.00314 [gr-qc]].

\bibitem{DeFelice:2024ops}
A.~De Felice and S.~Tsujikawa,
``Nonsingular black holes and spherically symmetric objects in nonlinear electrodynamics with a scalar field,''
Phys. Rev. D \textbf{111} (2025) no.6, 064051
[arXiv:2412.04754 [gr-qc]].

\bibitem{RBHLE}
D.~S.~J.~Cordeiro, E.~L.~B.~Junior, J.~T.~S.~S.~Junior, F.~S.~N.~Lobo, J.~A.~A.~Ramos, M.~E.~Rodrigues, D.~Rubiera-Garcia, L.~F.~D.~da Silva and H.~A.~Vieira,
``Regular black hole solutions with linear electrodynamics,'' [in preparation]


\bibitem{Cordeiro:2025ivw}
D.~S.~J.~Cordeiro, E.~L.~B.~Junior, J.~T.~S.~S.~Junior, F.~S.~N.~Lobo, J.~A.~A.~Ramos, M.~E.~Rodrigues, L.~F.~D.~da Silva and H.~A.~Vieira,
``Black bounce solutions via nonminimal scalar-electrodynamic couplings,''
[arXiv:2509.24053 [gr-qc]].



\bibitem{Lessa:2024erf}
L.~A.~Lessa and G.~J.~Olmo,
JCAP \textbf{03} (2025), 019
doi:10.1088/1475-7516/2025/03/019
[arXiv:2412.05378 [gr-qc]].

\bibitem{Hawking:1973uf}
S.~W.~Hawking and G.~F.~R.~Ellis,
``The Large Scale Structure of Space-Time,''
Cambridge University Press, 2023,
ISBN 978-1-009-25316-1, 978-1-009-25315-4, 978-0-521-20016-5, 978-0-521-09906-6, 978-0-511-82630-6, 978-0-521-09906-6
doi:10.1017/9781009253161

\bibitem{Israel:1966rt}
W.~Israel,
``Singular hypersurfaces and thin shells in general relativity,''
Nuovo Cim. B \textbf{44S10} (1966), 1
[erratum: Nuovo Cim. B \textbf{48} (1967), 463]
doi:10.1007/BF02710419

\bibitem{Feng:2023vqp}
J.~C.~Feng,
``Smooth metrics can hide thin shells,''
Class. Quant. Grav. \textbf{40} (2023) no.19, 197002
doi:10.1088/1361-6382/acf2de
[arXiv:2308.11885 [gr-qc]].

\bibitem{Baines:2023xje}
J.~Baines, R.~Gaur and M.~Visser,
``Defect Wormholes Are Defective,''
Universe \textbf{9} (2023) no.10, 452
doi:10.3390/universe9100452
[arXiv:2308.16624 [gr-qc]].

\end{thebibliography}
\end{document}